\documentclass[aps,prb,twocolumn,superscriptaddress]{revtex4-1}
\usepackage{amsmath,amsthm,amssymb}
\usepackage{graphicx}
\usepackage{amsmath,bm}
\usepackage{color,ulem}
\usepackage{ifthen}

\newcount \commentout
\newcommand{\1}{\mbox{1}\hspace{-0.25em}\mbox{l}}

\newcommand{\mat}[1]{\left( \begin{matrix} #1 \end{matrix} \right)}

\newlength{\figwidth}
\newlength{\figlarge}
\begin{document}
\title{
Exceptional rings protected by emergent symmetry for mechanical systems
}
\author{Tsuneya Yoshida}
\affiliation{Department of Physics, University of Tsukuba, Ibaraki 305-8571, Japan}
\author{Yasuhiro Hatsugai}
\affiliation{Department of Physics, University of Tsukuba, Ibaraki 305-8571, Japan}
\date{\today}
\begin{abstract}
We propose mechanical systems, described by Newton's equation of motion, as suited platforms for symmetry protection of non-Hermitian topological degeneracies.
We point out that systems possess emergent symmetry, which is a unique properties of mechanical systems. Because of the emergent symmetry, in contrast to other systems, fine-tuning of parameters (e.g., gain and loss) is not required to preserve the symmetry protecting exceptional rings in two dimensions.
The presence of symmetry-protected exceptional rings (SPERs) in two dimensions is numerically demonstrated for a mechanical graphene with friction.
Furthermore, classification of symmetry-protected non-Hermitian degeneracies is addressed by taking into account the above special characteristics of mechanical systems.
\end{abstract}
\pacs{
***
} 
\maketitle

\section{Introduction}
After the discovery of a topological insulator for a quantum well of HgTe/CdTe~\cite{Kane_Z2TI_PRL05,HgTe_Bernevig06,Konig_QSHE2007}, topological phenomena have been analyzed extensively~\cite{TI_review_Hasan10,TI_review_Qi10}.
Interestingly, topological phenomena can be observed even for classical systems. 
For instance, topologically protected excitations are reported for classical systems described by the Newton's low~\cite{Kane_NatPhys13,Kariyado_SR15,Suesstrunk_Mech-class_PNAS16}. 
Other representative examples are systems described by the Maxwell's equations, e.g., photonic crystals~\cite{Haldane_chiralPHC_PRL08,Raghu_chiralPHC_PRA08,Wang_chiralPHC_Nature09} and electric circuits~\cite{Ningyuan_Topoelecircit_PRX15,Victor_Topoelecircit_PRL15,Lee_Topoelecircit_CommPhys18}. 
In particular, chiral edge states~\cite{Thouless_PRL1982,Halperin_PRB82,Hatsugai_PRL93} realized for photonic crystals~\cite{Wang_chiralPHC_Nature09} provide the basis for novel devises (e.g, directional filters~\cite{Fu_chiralPHCapp_AppPhys10}), which enhances the significance of classical topological systems. 
The above topological phenomena both for quantum and classical systems are mathematically described as an eigenvalue problem of an Hermitian matrix.

Recently, theoretical and experimental efforts have opened a new field of research, non-Hermitian topological systems, where the eigenvalue problem becomes non-Hermitian~\cite{Hatano_PRL96,CMBender_PRL98,Fukui_nH_PRB98}. 
The non-Hermitian topological phenomena have been reported for a variety of systems, such as open quantum systems~\cite{TELeePRL16_Half_quantized,YXuPRL17_exceptional_ring,ZPGong_PRL17,HShen2017_non-Hermi,Carlstrom_nHknot_PRA18}, photonic crystals~\cite{Guo_nHExp_PRL09,Ruter_nHExp_NatPhys10,Regensburger_nHExp_Nat12,Zhen_AcciEP_Nat15,Hassan_EP_PRL17,Zhou_ObEP_Science18}, and correlated systems in equilibrium etc~\cite{VKozii_nH_arXiv17,Yoshida_EP_DMFT_PRB18,HShen2018quantum_osci}.
In this field, as well as the bulk-edge correspondence under the non-Hermitian skin effect~\cite{SYao_nHSkin-1D_PRL18,SYao_nHSkin-2D_PRL18,KFlore_nHSkin_PRL18,EElizabet_PRBnHSkinHOTI_PRB19}, the interplay between symmetry and the non-Hermiticity has been addressed as a central issue.
In particular, the non-Hermiticity unifies~\cite{Gong_class_PRX18,KKawabata_TopoUni_NatComm19}/ramifies~\cite{Kawabata_gapped_class_arXiv19} the symmetry classes.
Furthermore, recent theoretical studies~\cite{Budich_SPERs_PRB19,Okugawa_SPERs_PRB19,Yoshida_SPERs_PRB19} have revealed that symmetry enriches the topology of exceptional points which arise from defectiveness of the Hamiltonian~\cite{TKato_EP_book1966,HShen2017_non-Hermi} (i.e., the breakdown of its diagonalizability). These novel non-Hermitian degeneracies are referred to as symmetry-protected exceptional rings (SPERs) for two-dimensional systems and symmetry-protected exceptional surfaces (SPESs) for three-dimensional systems.

So far, open quantum systems~\cite{TELeePRL16_Half_quantized,YXuPRL17_exceptional_ring,ZPGong_PRL17,HShen2017_non-Hermi,Carlstrom_nHknot_PRA18} and photonic systems~\cite{Guo_nHExp_PRL09,Ruter_nHExp_NatPhys10,Regensburger_nHExp_Nat12,Zhen_AcciEP_Nat15,Hassan_EP_PRL17,Zhou_ObEP_Science18} have mainly been analyzed as the experimental platforms of the non-Hermitian topological physics.
In these systems, fine-tuning is required in experiments to analyze the interplay of symmetry and non-Hermiticity, such as emergence of the aforementioned symmetry-protected non-Hermitian degeneracies. 
For example, to preserve parity-time symmetry ($PT$-symmetry) in photonic systems, one need to tune gain and loss as well as coupling between sites~\cite{Takata_pSSH_PRL18,Takata_pSSH_pcomm}.
Therefore, for experimental realizations of symmetry-protected non-Hermitian degeneracies, it is important to find a system where relevant symmetry is preserved without fine-tuning.

We here elucidate that without fine-tuning, mechanical systems host symmetry-protected non-Hermitian topological degeneracies because of emergent symmetry. 
Specifically, we demonstrate the presence of SPERs with extended chiral symmetry~\cite{Yoshida_SPERs_PRB19} for a mechanical graphene with homogeneous fiction. 
Interestingly, in such a system, the zero-th Chern number characterizing the SPERs can be obtained by experimental observable quantities. 
Furthermore, we also carried out topological classification of symmetry-protected non-Hermitian degeneracies by taking into account the special characteristics of mechanical systems. The obtained result elucidates the robustness of SPERs in the mechanical graphene; SPERs survive even when the system is rotated because of the presence of $CP$-symmetry, while SPERs vanish further breaking the inversion symmetry with inhomogeneous gravitational potentials.

The rest of this paper is organized as follows.
In Sec. II, we elucidate that mechanical systems with friction hosts symmetry-protected non-Hermitian degeneracies with emergent symmetry by recasting the equation of motion as an eigenvalue problem of a non-Hermitian matrix.
In Sec. III, we demonstrate the emergence of SPERs with extended chiral symmetry for the mechanical graphene with homogeneous friction.
In Sec. IV, we address topological classification of symmetry-protected non-Hermitian topological degeneracies by taking into account the special characteristics of mechanical systems.

\section{
Symmetry-protected non-Hermitian degeneracies with emergent symmetry
}
\label{sec: generic_arg_of_mech}
Firstly, we show that a wide variety of mechanical systems host symmetry-protected non-Hermitian degeneracies, such as SPERs in two dimensions.
Consider a mechanical system with friction where internal forces (e.g., the Coriolis force) and the Lorentz force are absent.
In this case, the system preserves emergent symmetry~(\ref{eq: chiral}) which results in the above symmetry-protected non-Hermitian degeneracies.

In the following, we discuss the symmetry of the systems and discuss the symmetry-protection of the non-Hermitian degeneracies.
\subsection{
Matrix form of equation of motion
}
\label{sec: mapping EOM}

Let us consider a mechanical system, such as a coupled vibration system. 
Then, the equation of motion is given by
\begin{subequations}
\begin{eqnarray}
\label{eq: EOM_gene}
\ddot{u}^\mu_{\bm{k}} &=& -D^{\mu\nu}(\bm{k}) u^\nu_{\bm{k}} +\Gamma^{\mu\nu}_0(\bm{k}) \dot{u}^\nu_{\bm{k}},
\end{eqnarray}
\end{subequations}
where $u^\mu_{\bm{k}}$ denotes the Fourier transformed displacement ($\mu,\nu=x,y$). $\dot{u}^\mu_{\bm{k}}:=\frac{d u^\mu_{\bm{k}} }{dt}$.
Summation over repeating indices is assumed.
The first and second terms describes the potential force and the force proportional to the velocity.

In the matrix form, the above equation is rewritten as
\begin{subequations}
\label{eq: EOM gen mat}
\begin{eqnarray}
\dot{\bm{\phi}}_{\bm{k}}(t)
&=& 
M(\bm{k})
\bm{\phi}_{\bm{k}}(t), \\
M(\bm{k})
&=&
\mat{0 & \1 \\ -D(\bm{k}) & \Gamma_0(\bm{k})}_\rho, 
\end{eqnarray}
\end{subequations}
with 
$\bm{\phi}(t)=\mat{ \bm{u}_{\bm{k}} & \dot{\bm{u}}_{\bm{k}}}^T.$
The matrix $D(\bm{k})$ is Hermitian and positive definite, which guarantees that the system is mechanically stable.

Because the Hermitian matrix $D(\bm{k})$ is positive semi-definite, we can introduce a Hermitian matrix $Q$ with $D(\bm{k})=Q^2(\bm{k})$ (for more details, see Appendix~\ref{sec: symm Q(k)}).
Thus, the equation of motion~(\ref{eq: EOM gen mat}) is rewritten as~\cite{Kane_NatPhys13,Suesstrunk_Mech-class_PNAS16}
\begin{subequations}
\label{eq: BdG gen}
\begin{eqnarray}
i\partial_t\bm{\psi}(t) &=& H\bm{\psi}(t), \\
H
 &=& 
\left(
\begin{array}{cc}
0 &  Q(\bm{k})\\
Q(\bm{k}) & i\Gamma_0(\bm{k})
\end{array}
\right)_\rho,
\end{eqnarray}
with
\begin{eqnarray}
\bm{\psi}(t)
&=& 
\left(
\begin{array}{cc}
Q(\bm{k}) &  0\\
0 & i\1
\end{array}
\right)_\rho
\bm{\phi}(t).
\end{eqnarray}
\end{subequations}
The matrices $H$ and $M$ in Eq.~(\ref{eq: EOM gen mat}a) give the same eigenvalues (i.e., same frequency).
Therefore, the dynamics of the system is described by the matrix $H$ which is Hermitian in the absence of dissipation. 
Because Eq.~(\ref{eq: BdG gen}) is mathematically identical to the Schr\"odinger equation, we refer to the matrix $H$ as Hamiltonian in the following.
When $i\Gamma_0(\bm{k})$ is a Hermitian, the system conserve the energy. Otherwise the system is dissipative.

\subsection{
Emergent symmetry
}
\label{sec: symm EOM}
Remarkably, mechanical systems possess emergent symmetry; in contrast to other platforms of non-Hermitian physics (e.g., quantum systems or photonic crystals), the Hamiltonian $H$ describing mechanical systems preserves extended chiral symmetry~(\ref{eq: chiral}) regardless of its details. 
Furthermore, the system preserves $CP$-symmetry~(\ref{eq: CP}), provided that the inversion symmetry is present.
In the following, we see the details.

(i) \textit{Emergent symmetry--} 
Any system, whose matrix $\Gamma_0(\bm{k})$ purely describes loss of energy, possesses the following emergent symmetry (i.e., the extended chiral symmetry)
\begin{eqnarray}
\label{eq: chiral}
\rho_3 H^\dagger(\bm{k}) \rho_3 &=&-H(\bm{k}),
\end{eqnarray}
where $\rho$'s are Pauli matrices. 
Mathematically, the above emergent symmetry is preserved if $i\Gamma_0(\bm{k})$ is anti-Hermitian. Introducing the Coliofi force or the Lorentz force breaks the extended chiral symmetry.

(ii) \textit{$CP$-symmetry--} 
When the system is inversion symmetric, the $CP$-symmetry is preserved. 
To see this, we first mention particle-hole symmetry of mechanical systems.
Particle-hole symmetry is preserved for any $D$ and $\Gamma_0$ rewritten as real matrices in the real-space.
Namely, when the matrixes satisfy $D^*(\bm{k})=D(-\bm{k})$ and $\Gamma^*_0(\bm{k})=\Gamma_0(-\bm{k})$, the system preserves the particle-hole symmetry
\begin{eqnarray}
\label{eq: PH}
\rho_3 H^*(\bm{k}) \rho_3 &=&-H(-\bm{k}).
\end{eqnarray}
Thus, if the system is inversion symmetric [i.e, $U_{I}D(\bm{k})U _{I}=D(-\bm{k})$ and $U_{I}\Gamma_0(\bm{k})U _{I}=\Gamma_0(-\bm{k})$ with $U^2_{I}=\1$ hold], the Hamiltonian preserves the $CP$-symmetry
\begin{eqnarray}
\label{eq: CP}
U_{CP}H^*(\bm{k}) U^\dagger_{CP} &=&-H(\bm{k}),
\end{eqnarray}
with $U_{CP}=U_{I}\otimes\rho_3$.
$CP$-symmetry may exist even when the Coriolis force is present. However, to preserve the $CP$-symmetry, fine-tuning is necessary.

We note that the matrix $Q(\bm{k})$ inherits symmetry of the matrix $D(\bm{k})$ (for more details see Appendix~\ref{sec: symm Q(k)}).

\subsection{
Symmetry-protected non-Hermitian degeneracies with extended chiral symmetry
}
\label{sec: protection chiral}
Here, we show that the extended chiral symmetry~(\ref{eq: chiral}) results in symmetry-protected non-Hermitian topological degeneracies~\cite{Yoshida_SPERs_PRB19} by analyzing the Hamiltonian around a band touching point. 
The topological characterization of these degeneracies is discussed in Sec.~\ref{sec: mech_graph_simulation}.

Firstly, we note that in the presence of the extended chiral symmetry, the energy eigenvalues form a pair $(E,-E^*)$ or take pure imaginary values ($E\in i\mathbb{R}$) (see Appendix~\ref{sec: eivenval_chiral_app}).
This fact indicates that each pair $(E,-E^*)$ cannot split without going through a band touching.

Let us consider a one-dimensional line in the BZ where a pair of bands splits via a band touching point. As we see below, the Hamiltonian is defective at this band touching point.
Around but not at the band touching point, we first project the Hilbert space to the space spanned by the pair of bands. Then the Hamiltonian and the operator of extended chiral symmetry are reduced into $2\times 2$ matrices.
Here, the chral operator is written as $U_{\Gamma}=\bm{n}\cdot \bm{\sigma}$~\cite{footnote_2x2chiral} with a unit vector $\bm{n}\in \mathbb{R}^3$.
Therefore, without loss of generality, the Hamiltonian and the operator of extended chiral symmetry are written as
\begin{subequations}
\begin{eqnarray}
     H &=& id_0\sigma'_0 + (\bm{b}+i\bm{d})\cdot\bm{\sigma'}, \\
 U_{\Gamma} &=&\sigma_3,
\end{eqnarray}
with $\bm{b}=(b_1,b_2,0)$ and $\bm{d}=(0,0,d_3)$. 
The eigenvalues of the Hamiltonian $H$ are written as
\begin{eqnarray}
 E &=& id_0 \pm \sqrt{ b^2-d^2}.
\end{eqnarray}
\end{subequations}
Therefore, at the band touching point satisfying $b=d$, the Hamiltonian is written as
$
H=\left(
\begin{array}{cc}
id_0 & b \\
0 & id_0
\end{array}
\right)
$
with a proper choice of the basis.
This result indicates that the Hamiltonian is defective~\cite{footnote_ERchiral_b=d=0}.

We note that $CP$-symmetry also results in symmetry-protected non-Hermitian degeneracies, which can be seen in a similar way as the above case (see Appendix.~\ref{sec: protection cp}).

\section{
SPERs for a mechanical graphene with friction
}
We demonstrate the emergence of the SPERs for a mechanical graphene with friction (Fig.~\ref{fig: scketch of lattice}). 
\subsection{Model}
We first consider only potential forces. After that the Coriolis force and friction are introduced.

\begin{figure}[!h]
\begin{center}
\includegraphics[width=90mm,clip]{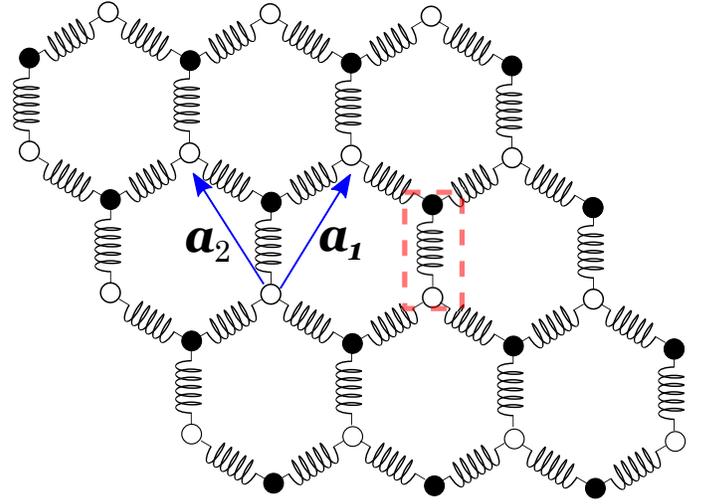}
\end{center}
\caption{(Color Online).
Sketch of the mechanical graphene where mass points are connected by springs.
The system is composed of A and B sublattices represented with white and black circles.
The unit cell is enclosed with the red dashed box.
$\bm{a}_1$ and $\bm{a}_2$ are the primitive translation vectors.
}
\label{fig: scketch of lattice}
\end{figure}

We consider that the spring-mass forms the honeycomb lattice whose unit cell is illustrated with red dashed box in Fig.~\ref{fig: scketch of lattice}. 
In this figure, primitive translation vectors $\bm{a}_1$ and $\bm{a}_2$ are shown with black arrows.
We also suppose that the mass points are trapped in dents of the floor, which introduces gravitational potentials.
When the displacement is sufficiently smaller than the lattice constant, the Lagrangian is represented as the following quadratic form,
\begin{subequations}
\label{eq: L of mech_graph}
\begin{eqnarray}
\mathcal{L}
 &=&
\frac{m}{2}
\sum_{\bm{k},\mu,\alpha}  \dot{u}^{\mu}_{\bm{k}\alpha} \cdot \dot{u}^{\mu}_{-\bm{k}\alpha} - \frac{1}{2}\sum_{\bm{k}}  D^{\mu\nu}_{\alpha\beta}(\bm{k}) u^\mu_{\bm{k},\alpha} u^\nu_{-\bm{k},\beta}, \nonumber \\
\end{eqnarray}
with $D^{\mu\nu}_{\alpha\beta}(\bm{k})=D_{\mu\alpha;\nu\beta}(\bm{k})$ and
\begin{eqnarray}
D(\bm{k})
 &=& 
 3\kappa(1-\frac{\eta}{2})\1+
 \mat{  \kappa^0_A & D_{AB}(\bm{k})  \\  D^\dagger_{AB} (\bm{k})  & \kappa^0_B}, \nonumber \\
&& \\
D_{AB}(\bm{k}) &=& 
\kappa( \gamma_3+ \gamma_1 e^{-i\bm{k}\cdot \bm{a}_1}+ \gamma_2 e^{-i\bm{k}\cdot \bm{a}_2 }), \\
\gamma_1&:=& (1-\eta) \mat{ 1 & 0 \\  0 & 1} + \eta \mat{ \frac{3}{4} & \frac{\sqrt{3}}{4}  \\  \frac{\sqrt{3}}{4}  & \frac{1}{4} } , \\
\gamma_2&:=& (1-\eta) \mat{ 1 & 0 \\  0 & 1} + \eta \mat{ \frac{3}{4} & -\frac{\sqrt{3}}{4}  \\  -\frac{\sqrt{3}}{4}  & \frac{1}{4} } , \\
\gamma_3&:=& (1-\eta) \mat{ 1 & 0 \\  0 & 1} + \eta \mat{ 0 & 0  \\  0  & 1 } . 
\end{eqnarray}
\end{subequations}
Here, $m$ denotes the mass, and $\kappa$ describes the potential force of springs.
$\kappa^0_\alpha$ denotes the strength of the gravitational force for sublattice $\alpha$ ($\alpha=A,B$).
$u^\mu_{\bm{k}\alpha}$ denotes the Fourier transformed displacement of the mass point at $\alpha$-sublattice.
The $\mu(=x,y)$ denotes the $\mu$ component.
$\eta:=l_0/R$ where $l_0$ denotes natural length of the spring and $R$ denotes lattice constant ($R=1$).

Taking variational derivative of the Lagrangian, we obtain the equation of motion
\begin{eqnarray}
\label{eq: EOM_nondiss}
\ddot{u}^\mu_{\bm{k}\alpha} &=& -D^{\mu\nu}_{\alpha\beta}(\bm{k}) u^\nu_{\bm{k}\beta}.
\end{eqnarray}

Now, let us introduce the frictional force and the Coriolis force which are proportional to velocity, the equation of motion is written as
\begin{subequations}
\begin{eqnarray}
\label{eq: EOM_diss}
\ddot{u}^\mu_{\bm{k}\alpha} &=& -D^{\mu\nu}_{\alpha\beta}(\bm{k}) u^\nu_{\bm{k}\beta} +\Gamma^{\mu\nu}_{0\alpha\beta}(\bm{k}) \dot{u}^\nu_{\bm{k}\beta}, \\
\Gamma^{\mu\nu}_{0\alpha\beta} &=&-b_\alpha \delta^{\mu\nu}\delta_{\alpha\beta}+\Omega_0 \epsilon^{\mu\nu}\delta_{\alpha\beta},
\end{eqnarray}
\end{subequations}
with $\Omega_0\in\mathbb{R}$ and $b_\alpha \geq 0$. 
Here, summation over repeating indices is assumed. $\epsilon^{\mu\nu}$ is an anti-symmetric matrix with $\epsilon^{xy}=1$.
The term proportional to $b_\alpha$ describes the frictional force, and the term proportional to $\Omega_0$ describes the Coriolis force (i.e., $\Omega_0$ denotes the angular velocity).

We note that the above equations is rewritten in the form of Eq.~(\ref{eq: BdG gen}) with 
$
\bm{\phi}=(\bm{u}_{\bm{k}},\dot{\bm{u}}_{\bm{k}})^T
$ 
and 
$
\bm{u}_{\bm{k}}
=
\mat{   u^x_{\bm{k}A} & u^y_{\bm{k}A} & u^x_{\bm{k}B} & u^y_{\bm{k}B} }^T
$. 

Here we briefly summarize the symmetry.
When the Coriolis force is absent, the system preserves the emergent symmetry~(\ref{eq: chiral}) regardless of the other details of the system.
Besides the emergent symmetry, $CP$-symmetry~(\ref{eq: CP}) is preserved because our system is inversion symmetric.
Due to the above symmetry, SPERs may emerge in the mechanical graphene.

\subsection{
Numerical demonstration
}
\label{sec: mech_graph_simulation}
We demonstrate that the mechanical graphene hosts SPERs with the emergent symmetry and $CP$-symmetry.
Specifically, after the results of the conservative system, we show the presence of SPERs with the emergent symmetry~(\ref{eq: chiral}) for the system with friction (Sec.~\ref{sec: mech_fric}). In Sec.~\ref{sec: rot_mech_fric}, we show that SPERs survive even in the absence of the emergent symmetry. This is because the system preserves $CP$-symmetry~(\ref{eq: CP}). In Sec.~\ref{sec: rot_mech_fric_w_inhomo}, we see that breaking both of the emergent symmetry and $CP$-symmetry changes the SPERs to exceptional points.

In the following, we discuss the data for $m=\kappa=1$.
\subsubsection{
System without friction
}
\label{sec: mech_graph_sim_conservative}
\begin{figure}[!h]
\begin{minipage}{1\hsize}
\begin{center}
\includegraphics[width=\hsize,clip]{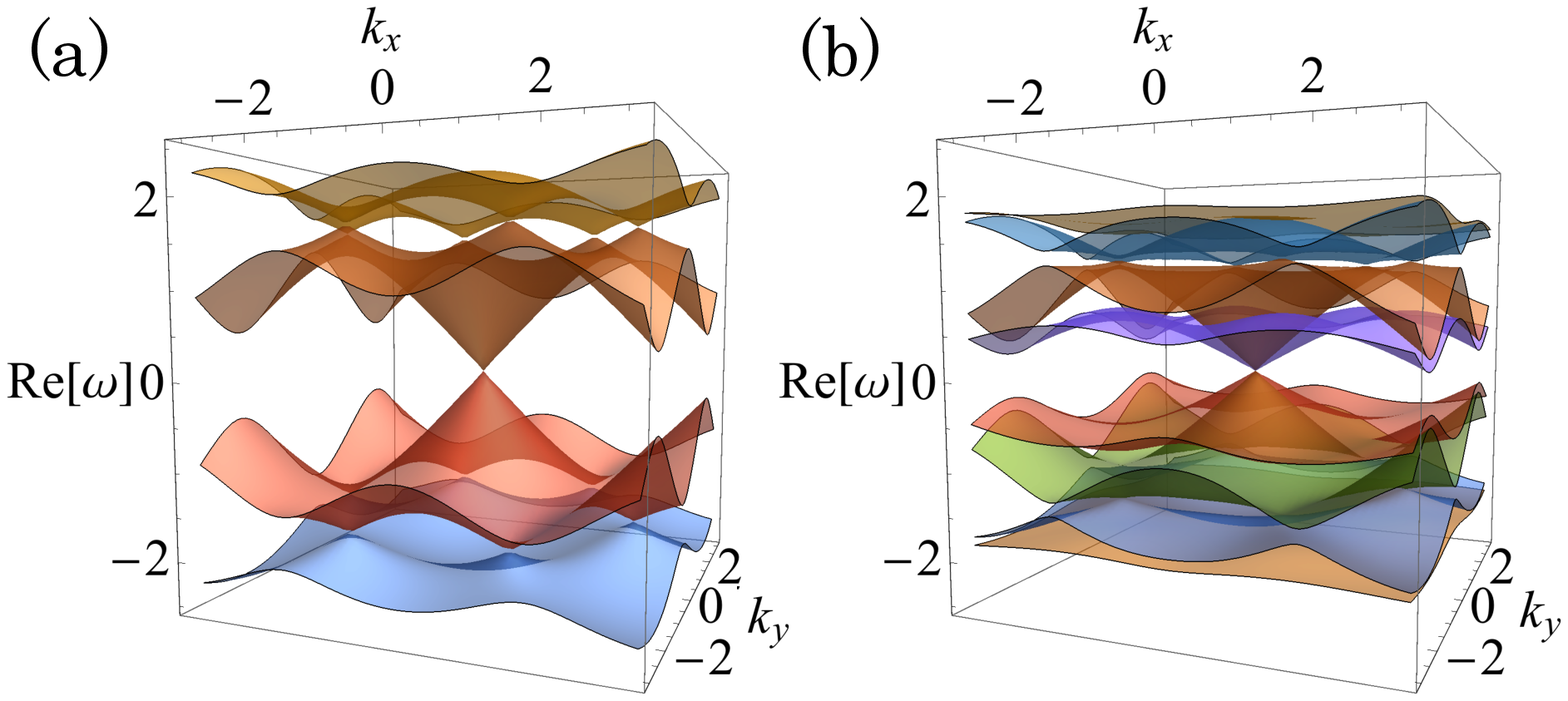}
\end{center}
\end{minipage}
\caption{(Color Online).
(a) [(b)] Energy dispersion $\omega(\bm{k})$ of the mechanical graphene for $(b,\eta)=(0,0)$ and [$(b,\eta)=(0,0.8)$].
}
\label{fig: En_nondissi}
\end{figure}

Let us start with the case of $b=0$ where the system is not dissipative.
The band structure for $\eta=0$ is plotted in Fig.~\ref{fig: En_nondissi} (a). 
In this case, the longitudinal and transverse waves are decoupled, which results in two-fold degeneracy of each band as shown in  Fig.~\ref{fig: En_nondissi} (a).
Increasing $\eta$ lifts the degeneracy [see Fig.~\ref{fig: En_nondissi} (b)]. For both cases of $\eta=0$ and $\eta=0.8$, one can observe the following two behaviors:
(i) the linear dispersion is observed around the $\Gamma$-point, which corresponds to the Nambu-Goldstone modes; (ii) the real-part of the energy eigenvalue is symmetric for $\mathrm{Re}[\omega]=0$, which arises from extended chiral symmetry and $CP$-symmetry [see Eqs.~(\ref{eq: chiral})~and~(\ref{eq: CP})].

\begin{figure}[!h]
\begin{minipage}{1\hsize}
\begin{center}
\includegraphics[width=\hsize,clip]{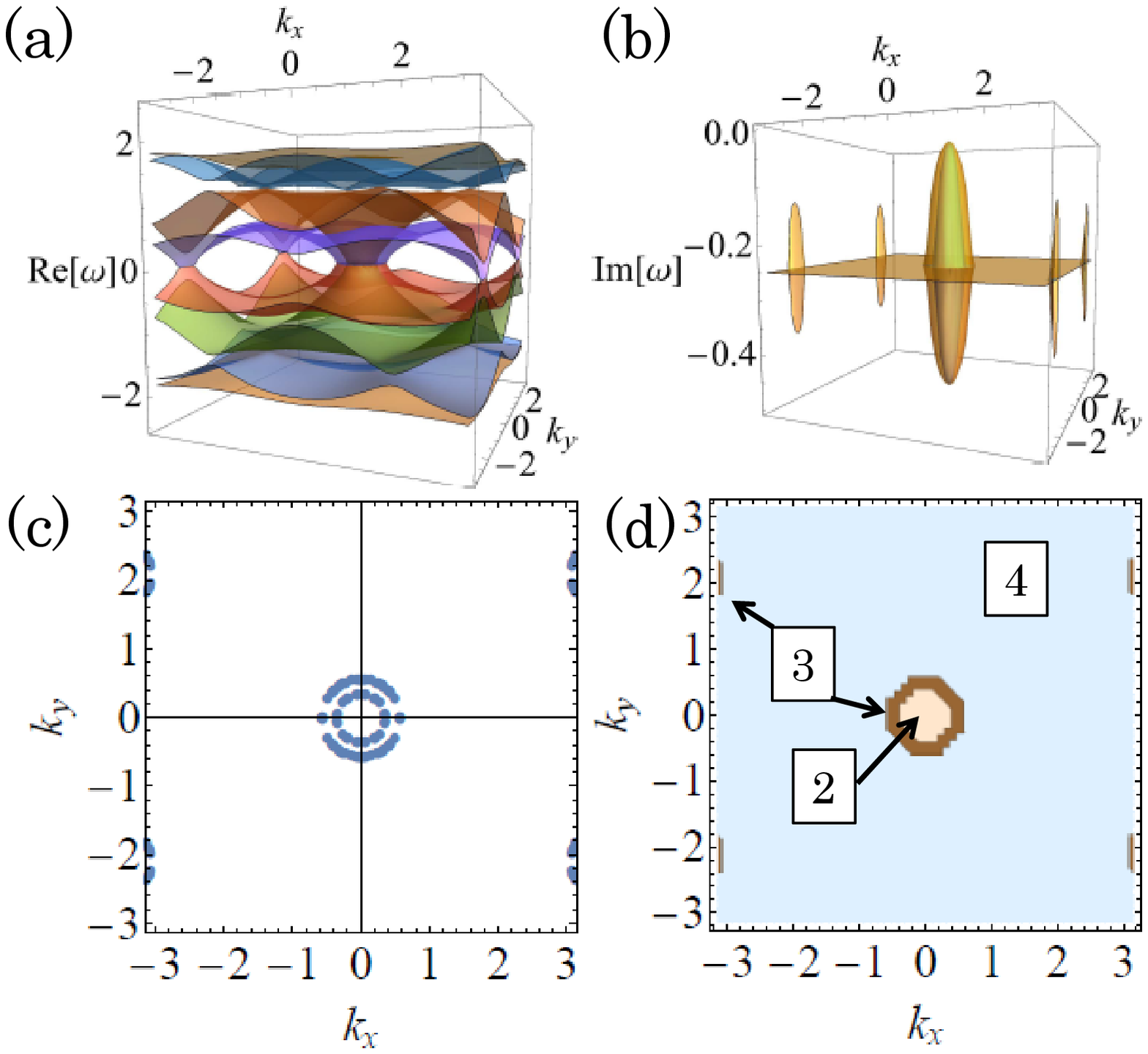}
\end{center}
\end{minipage}
\caption{(Color Online).
Numerical results of the mechanical graphene with friction. 
(a) [(b)] The real-part [imaginary-part] of the eigenvalues $\omega(\bm{k})$.
(c) Momentum points $\bm{k}_0$ satisfying $|\mathrm{det} U(\bm{k}_0)|<0.005$.
(d) The zero-th Chern number $N_{0\mathrm{Ch}}$ (i.e., the number of occupied bands for $-i H'\rho_3$ with $H'=H-\mathrm{tr}H/\mathrm{dim}H\1$) at each point of the BZ. 
For computation of $N_{0\mathrm{Ch}}$, we have used Eq.~(\ref{eq: simp_0Ch_mech}).
The data are obtained for $\Omega=\kappa_A=\kappa_B=0$ and $(b,\eta)=(0.5,0.8)$ with $b_A=b_b=b=0.5$.
}
\label{fig: En}
\end{figure}

\subsubsection{
System with friction
}
\label{sec: mech_fric}
Now, we move on to the system with friction by setting $b=0.5$. In this case, the eigenvalues $\omega$'s take complex values.
The real-part [the imaginary-part] of the energy eigenvalues for $\eta=0.8$ are plotted in Fig.~\ref{fig: En}(a) [(b)], respectively.
Indeed, the Hamiltonian becomes defective (i.e., the Hamiltonian is not diagonalizable) on this ring as we can see in Fig.~\ref{fig: En}(c).
This figure shows points in the BZ where determinant of the matrix $U(\bm{k})$ becomes zero. Here, $n$-th column of matrix $U(\bm{k})$ is the eigenvector of $M(\bm{k})$. 
Therefore, on the points plotted in Fig.~\ref{fig: En}(c), the Hamiltonian becomes defective.
We note that for some modes, the damping rate inside of the SPER becomes smaller than the one outside of the SPER [see Fig.~\ref{fig: En}(b)].
Therefore, the emergence of SPER results in the following dynamical behavior which may be observed in experiments.
A region emerges in the BZ for which the displacement damps slowly compared to the ones for the other region.

\begin{figure}[!h]
\begin{minipage}{1\hsize}
\begin{center}
\includegraphics[width=\hsize,clip]{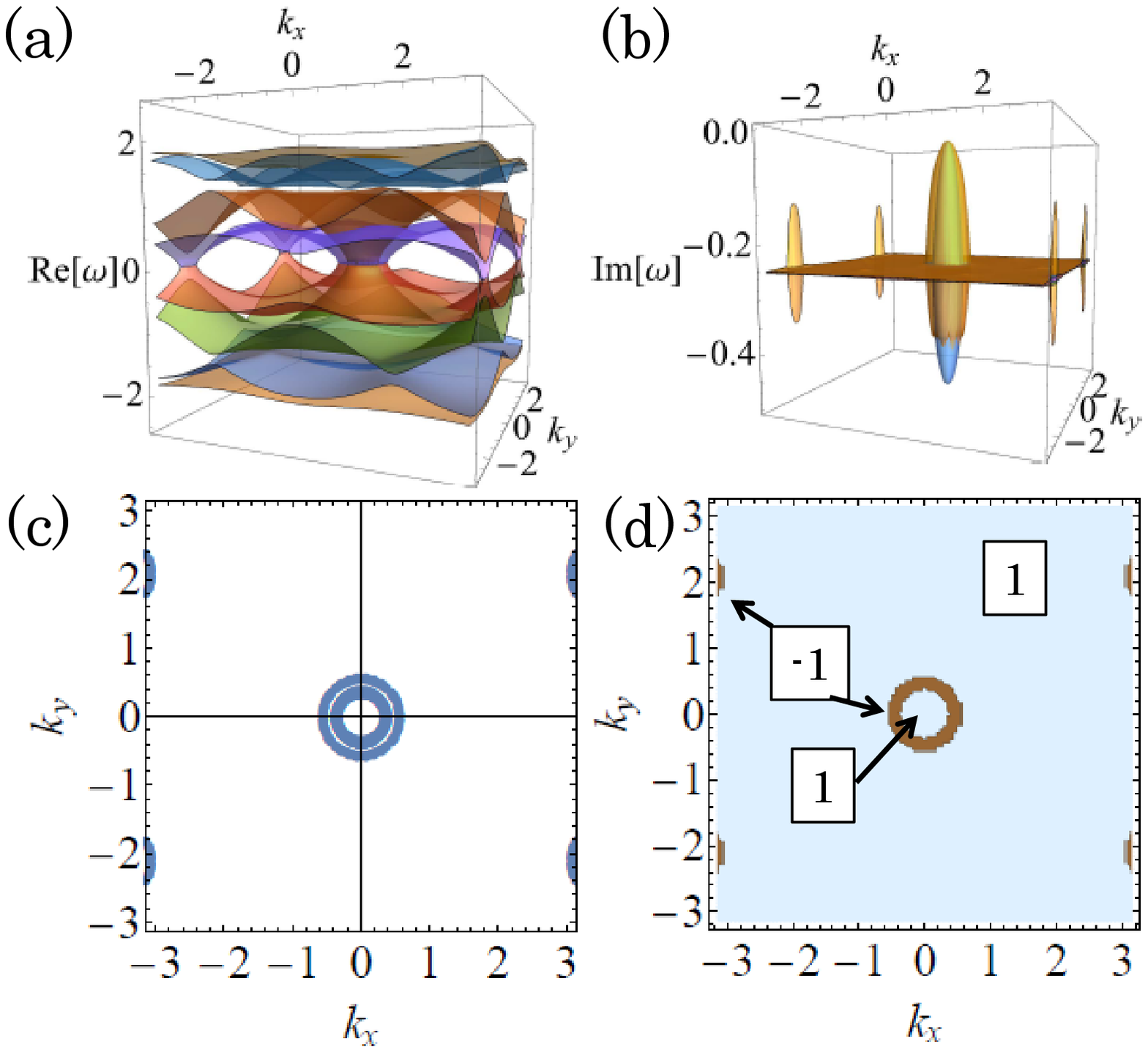}
\end{center}
\end{minipage}
\caption{(Color Online).
Numerical results of the rotating mechanical graphene with homogeneous friction.
(a) [(b)] The real-part [imaginary-part] of the eigenvalues $\omega(\bm{k})$.
(c) Momentum points $\bm{k}_0$ satisfying $|\mathrm{det} U(\bm{k}_0)|<0.005$.
(d) The $\mathbb{Z}_2$-invariant~(\ref{eq: Z2inv}) for each point of the BZ.
The data are obtained for $\kappa_A=\kappa_B=0$ and $(b,\Omega_0,\eta)=(0.5,0.05,0.8)$ with $b_A=b_b=b=0.5$.
}
\label{fig: En_rot}
\end{figure}

Here, we address the characterization of the exceptional points.
As pointed out in Ref.~\onlinecite{Yoshida_SPERs_PRB19}, SPERs are characterized with the zero-th Chern number (for the definition is in Appendix~\ref{sec: 0Ch_app}).
This fact is consistent with $\mathbb{Z}$-classification for class AIII with $\delta=1$ (see Sec.~\ref{sec: 10foldway_class}).
Figure~\ref{fig: En}(d) shows that the SPERs emerge at the boundary of two domains, having distinct values of $N_{0\mathrm{Ch}}$, which characterizes the SPERs.
Interestingly, when the frictional force is homogeneous, the zero-th Chen number $N_{0\mathrm{Ch}}$ can be simplified as
\begin{eqnarray}
\label{eq: simp_0Ch_mech}
 N_{0\mathrm{Ch}}(\bm{k})&=& \sum_{n=1,\cdots,\mathrm{dim}H} \Theta_>(2\omega_{0n}(\bm{k})-b),
\end{eqnarray}
when the frictional force is homogeneous.
$\Theta_>(x)$ takes $1$, $1/2$, and $0$ for $x>0$, $x=0$, and $x<0$, respectively. $\omega_{0n}$ ($n=1,\cdots,\mathrm{dim}H$) is the energy eigenvalues of the system without friction.
The detail of the derivation is summarized in Appendix~\ref{sec: 0Ch_app}.
Remarkably, the right hand side of Eq.~(\ref{eq: simp_0Ch_mech}) is composed only with experimental observables, indicating that the $N_{0\mathrm{Ch}}$ is measurable in experiments.

We consider that the zero-th Chern number and the presence of the region where the modes damp slowly are experimental signals of SPERs.
\subsubsection{
Rotating system with homogeneous friction
}
\label{sec: rot_mech_fric}
As seen in the Sec.~\ref{sec: protection chiral}, the robustness of the exceptional rings in this two-dimensional system arises from the extended chiral symmetry~(\ref{eq: chiral}). 
Here, we see that the SPERs in the mechanical graphene can survive even in the absence of the extended chiral symmetry.
In order to see this, we here analyze effects of the symmetry breaking. Firstly, we analyze the system by introducing the Coriolis force which breaks the extended chiral symmetry. 
Figs.~\ref{fig: En_rot}(a) and (b) show the real-part and the imaginary-part of the eigenvalues for the parameter set $(b_A,b_B,\Omega_0,\kappa_B,\eta)=(0.5, 0.5,0.05,0,0.8)$, respectively.
These figures show that band touching points form a ring. 
Fig.~\ref{fig: En_rot}(c) elucidates the points where the matrix $M(\bm{k})$ becomes defective. These figures indicate that the SPERs survive even in the absence of the extended chiral symmetry. In this case, the SPERs are protected by $CP$-symmetry~(\ref{eq: CP}) [see Appendix~\ref{sec: CP_transp_app}].

Classification results obtained in Sec.~\ref{sec: 10foldway_class} elucidate that the SPERs can be characterized by the $\mathbb{Z}_2$-invariant.
The $\mathbb{Z}_2$-invariant of the zero-dimensional systems is given by~\cite{Gong_class_PRX18}
\begin{eqnarray}
\label{eq: Z2inv}
 s(\bm{k})&=& \mathrm{sgn}[\mathrm{det}iH(\bm{k})],
\end{eqnarray}
where $\mathrm{sgn}(x)$ take 1 and $-1$ for $x>0$ and $x<0$.
In Fig.~\ref{fig: En_rot}(d), one can see that the $\mathbb{Z}_2$-invariant characterize the SPERs.

\begin{figure}[!h]
\begin{minipage}{1\hsize}
\begin{center}
\includegraphics[width=\hsize,clip]{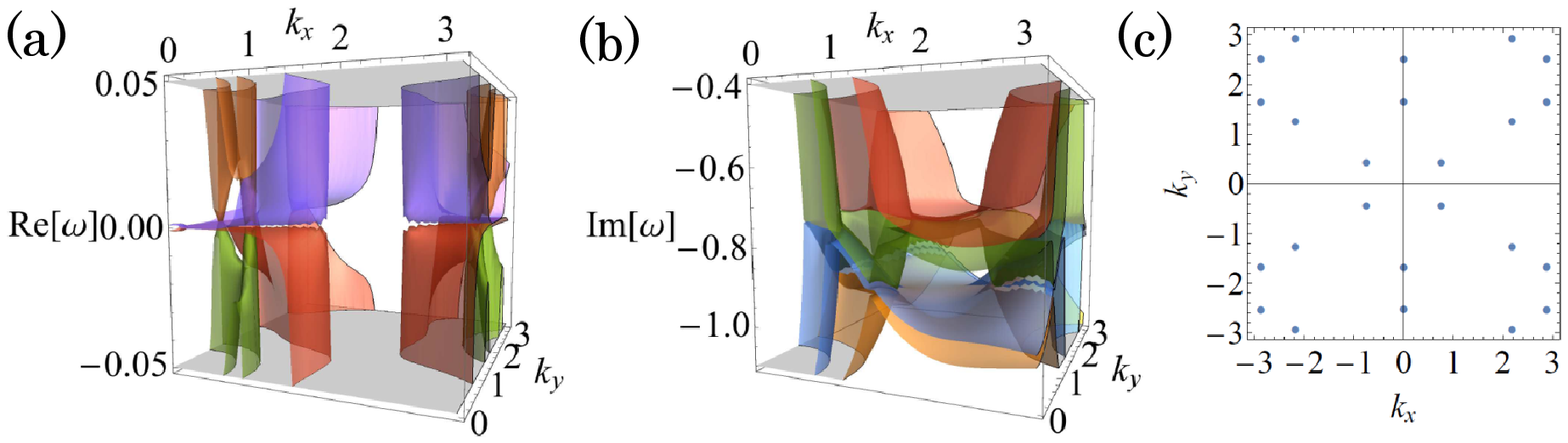}
\end{center}
\end{minipage}
\caption{(Color Online).
Numerical results of the rotating mechanical graphene with inhomogeneous friction and the gravitational potential.
The data are obtained for $(b_A,\Omega_0,\kappa_B,\eta)= (2,0.05,0.05,0.8)$ and $b_B=\kappa_A=0$.
(a) [(b)] The real-part [imaginary-part] of the eigenvalues $\omega(\bm{k})$.
(c) Momentum points $\bm{k}_0$ satisfying $|\mathrm{det} U(\bm{k}_0)|<0.005$.
}
\label{fig: En_break_inhomo}
\end{figure}

\subsubsection{
Rotating system with inhomogeneous friction and the gravitational potential
}
\label{sec: rot_mech_fric_w_inhomo}
Further breaking $CP$-symmetry changes the SPER to the exceptional points. Figure~\ref{fig: En_break_inhomo} shows data for the parameter set $(b_A,b_B,\Omega_0,\kappa_A,\kappa_B,\eta) = (2,0,0.05,0,0.05,0.8)$.
Figures~\ref{fig: En_break_inhomo}(a) and (b) show that band touching occurs at points rather than rings~\cite{footnote_bandtouch}. 
Correspondingly, the points, where the determinant of the matrix $U$ approaches to zero, does not form rings [see Fig.~\ref{fig: En_break_inhomo}(c)].
With breaking $CP$-symmetry but preserving extended chiral symmetry, we can observe the SPERs with extended chiral symmetry.
The above results indicate that exceptional points appear for mechanical systems when the systems have neither the extended chiral symmetry nor $CP$-symmetry.

\section{
Classification of the non-Hermitian degeneracies in mechanical systems
}
\label{sec: 10foldway_class}
So far, we have observed the existence of SPERs in the mechanical graphene due to emergent symmetry~(\ref{eq: chiral}), demonstrating that a wide range of mechanical systems host symmetry-protected non-Hermitian degeneracies.
In this section, for systematic understanding, we address the topological classification of symmetry-protected non-Hermitian degeneracies.

The above issue has recently been addressed for generic non-Hermitian matrices~\cite{Kawabata_gapless_class_arXiv19}. However, we would like to stress that the special characteristics, such as the emergent symmetry, needs to be take into account for systematic understanding of non-Hermitian degeneracies in mechanical systems, which is supported by the numerical results in Sec.~\ref{sec: mech_graph_simulation}.

\subsection{
Relevant symmetry
}
In Sec.~\ref{sec: symm EOM}, we have seen that mechanical systems may preserve the emergent symmetry~(\ref{eq: chiral}) and the $CP$-symmetry~(\ref{eq: CP}). 
Therefore, we address the topological classification for the following symmetry constraints:
\begin{subequations}
\label{eq: all symm}
\begin{eqnarray}
U_{TP} H^T(\bm{k}) U^\dagger_{TP} &=& H(\bm{k}), \\
U_{CP} H^*(\bm{k}) U^\dagger_{CP} &=& -H(\bm{k}), \\
U_{\Gamma} H^\dagger(\bm{k}) U^\dagger_{\Gamma} &=& -H(\bm{k}).
\end{eqnarray}
\end{subequations}
Eqs.~(\ref{eq: all symm}b)~and~(\ref{eq: all symm}c) are generic forms of $CP$-symmetry and extended chiral symmetry discussed in Sec.~\ref{sec: symm EOM}.
The first condition is obtained by the product of Eqs.~(\ref{eq: all symm}b)~and~(\ref{eq: all symm}c).

We note that another type of $CP$-symmetry $U_{CP} H^T(\bm{k}) U^\dagger_{CP} = -H(\bm{k})$ is incompatible with homogeneous friction (for more details see Appendix~\ref{sec: CP_transp_app}). Therefore, we only consider the symmetry shown in Eq.~(\ref{eq: all symm}).

\subsection{
Brief description of classification scheme
}
\label{sec: class_scheme_nonH}
\subsubsection{
Band gap for non-Hermitian systems
}
As pointed out in Ref.~\onlinecite{Kawabata_gapped_class_arXiv19}, there are two ways to define the band gap for the non-Hermitian systems because the eigenvalues take complex values.
The first one is the point gap~\cite{Gong_class_PRX18} and the second one is the line gap~\cite{HShen2017_non-Hermi}. 
The results obtained in Sec.~\ref{sec: mech_graph_simulation} indicate that adopting the point gap captures the SPERs and the SPESs, which we see below.

Firstly, we define that a system is gapped when its Hamiltonian $H(\bm{k})$ satisfies $\mathrm{det}H(\bm{k})\neq 0$. Secondly, consider two subsystems in the BZ whose point gap is finite. If the two Hamiltonians describing these subsystems are topologically distinct, then, the gap needs to close ($\mathrm{det}H=0$) in a region separating the two subsystems, which corresponds to the non-Hermitian degeneracies.

\subsubsection{
Classification scheme
}
The classification of $d_{EP}$-dimensional non-Hermitian degeneracies (i.e., gapless nodes) in $d$ spatial dimensions is accomplished by classifying $\delta-1$ dimensional gapped Hamiltonian with $\delta=d-d_{EP}$.

The strategy of the classification is mapping the non-Hermitian Hamiltonian to a Hermitian matrix whose classification is well-established.
Such mapping is accomplished by defining
\begin{eqnarray}
\tilde{H}(\bm{k}) &:=&
\left(
\begin{array}{cc}
 0 & H(\bm{k}) \\
H^\dagger(\bm{k}) & 0
\end{array}
\right)_\chi.
\end{eqnarray}
The condition $\mathrm{det}H=0$ is equivalent to $\mathrm{det}\tilde{H}=0$, indicating that the gap-closing of the non-Hermitian Hamiltonian $H$ can be captured with Hermitian Hamiltonian $\tilde{H}$.

Now we discuss the symmetry constraints on $\tilde{H}$.
The $TP$-symmetry~(\ref{eq: all symm}a) is rewritten as
\begin{subequations}
\label{eq: all symm tilde}
\begin{eqnarray}
\widetilde{TP} \tilde{H}(\bm{k}) \widetilde{TP}^{-1} &=& \tilde{H}(\bm{k}),
\end{eqnarray}
with $\widetilde{TP} =U_{TP}\otimes \chi_1\mathcal{K}$.
The $CP$-symmetry~(\ref{eq: all symm}a) is rewritten as
\begin{eqnarray}
\widetilde{CP} \tilde{H}(\bm{k}) \widetilde{CP}^{-1} &=& -\tilde{H}(\bm{k}),
\end{eqnarray}
with $\widetilde{CP}=U_{CP}\otimes\chi_0\mathcal{K}$.
The extended chiral symmetry~(\ref{eq: all symm}a) is rewritten as
\begin{eqnarray}
\tilde{\Gamma} \tilde{H}(\bm{k}) \tilde{\Gamma}^{-1} &=& -\tilde{H}(\bm{k}),
\end{eqnarray}
with $\tilde{\Gamma}=U_{\Gamma}\otimes\chi_1$.
\end{subequations}
Here, the operator $\mathcal{K}$ takes complex conjugation. $\chi$'s are the Pauli matrices acting on the extended Hilbert space. 
In addition to the above symmetry, the Hermitian Hamiltonian $\tilde{H}$ always preserves the chiral symmetry $\tilde{\Sigma}:=\1\otimes\chi_3$.

Therefore, the topological classification of non-Hermitian degeneracies is reduced to the classification of the Hermitian Hamiltonian $\tilde{H}$ for ten-fold way symmetry classes, which are called AZ$+\mathcal{I}$ classes~\cite{Bzdusek_AZ+I_PRB17}, in the presence of additional chiral symmetry $\Sigma=\1\otimes\chi_3$.
The latter problem can be solved by approach based on the Clifford algebra.
The detailed calculations are shown in Appendix~\ref{sec: class_Hermi_Sig_app}.

Here, a comment is in order concerning the special characteristics of the mechanical systems.
For symmetry class CII, C, and CI, additional $CP$-symmetry [$CP'=U_{CP'}\mathcal{K}$ and $(CP')^2=1$] should be taken into account as well as $CP$-symmetry with $CP=U_{CP}\mathcal{K}$ and $(CP)^2=-1$. 
This is because particle-hole symmetry~(\ref{eq: PH}) is generically preserved for mechanical systems; Eq.~(\ref{eq: PH}) is satisfied for any $D$ and $\Gamma_0$ rewritten as real matrices in the real-space.
Thus, we need to carry out the classification for the corresponding Hermitian systems by taking into account the additional chiral symmetry $\Sigma$ and $CP'$-symmetry, which is discussed in Appendix~\ref{sec: class_Hermi_Sig_CII_C_CI_app}.

\subsection{
Classification results
}
Tables~\ref{table: class_table_mech}~and~\ref{table: class_table_mech_CII_CI} summarize the classification results of $d_{EP}$-dimensional non-Hermitian topological degeneracies in the $d$-dimensional BZ, specifying the relevant symmetry of the mechanical systems.
This classification results elucidate the presence/absence of a topological invariant for the $\delta-1$-dimensional subspace of the BZ; e.g., SPERs observed in Sec.~\ref{sec: mech_graph_simulation} correspond to the case of $\delta=1$ [$(d,d_{EP})=(2,1)$]. 

Table~\ref{table: class_table_mech} supports that a wide range of mechanical systems  host SPERs in two dimensions. This can be seen by noticing that the systems preserve the emergent symmetry~(\ref{eq: chiral}) when $\Gamma(\bm{k})$ describes only dissipative forces and that these systems belong to class AIII if further symmetry is absent.
Table~\ref{table: class_table_mech} also indicates that rotating systems with inversion symmetry host SPERs [see the result of class D for $\delta=1$ ].

Table~\ref{table: class_table_mech_CII_CI} shows that classification results depends on commutation or anti-commutation relation of the two operators, $CP$ and $CP'$, which is one of special characteristics for mechanical systems. This table indicates the existence of symmetry-protected non-Hermitian degeneracies for class $\mathrm{CII}_+$ which are characterized with a zero-dimensional topological invariant. We discuss this result by analyzing a $4\times 4$ matrix.

In the following, we systematically discuss the non-Hermitian degeneracies for the mechanical graphene by making use of the classification results (Tables~\ref{table: class_table_mech}). In addition, we discuss this result of class $\mathrm{CII}_+$ by analyzing a $4\times 4$ matrix.

\begin{table}[htb]
\begin{center}
\begin{tabular}{c c c c c c c c c c } \hline\hline
     & $TP$ & $CP$ & $\Gamma$         & homotopy                              & $\delta=1$           & $2$            &    $3$         & $4$            \\ \hline
A    & $0$  & $0$  & $0$              &     $\pi_0(C_{\delta})$               &    0                 & $\mathbb{Z}$   &    0           & $\mathbb{Z}$   \\ 
AIII & $0$  & $0$  & $1$              &     $\pi_0(C_{\delta-1})$             &  $\mathbb{Z}$        &    0           &  $\mathbb{Z}$  & 0              \\ \hline
AI   & $1$  & $0$  & $0$              &     $\pi_0(R_{\delta+6})$             &    0                 & $\mathbb{Z}$   & $\mathbb{Z}_2$ & $\mathbb{Z}_2$ \\ 
BDI  & $1$  & $1$  & $1$              &     $\pi_0(R_{\delta+7})$             &  $\mathbb{Z}$        & $\mathbb{Z}_2$ & $\mathbb{Z}_2$ &    0           \\ 
D    & $0$  & $1$  & $0$              &     $\pi_0(R_{\delta  })$             &  $\mathbb{Z}_2$      & $\mathbb{Z}_2$ &    0           & $\mathbb{Z}$   \\ 
DIII & $-1$ & $1$  & $1$              &     $\pi_0(R_{\delta+1})$             &  $\mathbb{Z}_2$      &    0           & $\mathbb{Z}$   &    0           \\ 
AII  & $-1$ & $0$  & $0$              &     $\pi_0(R_{\delta+2})$             &    0                 & $\mathbb{Z}$   &    0           &    0           \\ 
CII  & $-1$ & $-1$ & $1$              &     --                                &    --                &    --          &    --          &    --          \\ 
C    & $0$  & $-1$ & $0$              &     $\pi_0(C_{\delta})$               &    0                 & $\mathbb{Z}$   &    0           & $\mathbb{Z}$   \\ 
CI   & $1$  & $-1$ & $1$              &     --                                &    --                &    --          &    --          &    --          \\ \hline \hline
\end{tabular}
\end{center}
\caption{
Classification results for each case of codimension $\delta=d-d_{EP}$. Here we consider $d_{EP}$-dimensional gapless node for $d$ spatial dimensions.
``$0$" in the second, the third, and the fourth columns denotes that the corresponding symmetry is absent. The $\pm 1$ in the second [the third] column represents the sign of $(TP)^2=\pm 1$ [$(CP)^2=\pm 1$], respectively. 
From sixth to ninth columns, the classification results are summarized where $\mathbb{Z}$ or $\mathbb{Z}_2$ means the presence of topological phases with the corresponding topological invariant. ``0" in these columns indicates the absence of topological phases.
For symmetry classes, CII, C, and CI, systems preserve $CP'$ symmetry with $(CP')^2=1$ as well as $CP$-symmetry with $(CP)^2=-1$. This additional symmetry changes the classification results for these three symmetry classes.
Thus, the results for CII and CI are shown in Table~\ref{table: class_table_mech_CII_CI}.
The classification results for $d$-dimensional gapped systems are obtained for $\delta=d+1$.
}
\label{table: class_table_mech}
\end{table}

\begin{table}
\begin{center}
\begin{tabular}{c c c c c c c c c } \hline\hline
                     & $TP$ & $CP$ & $\Gamma$         & homotopy                   & $\delta=1$      & 2               &    3            & 4               \\ \hline
$\mathrm{CII}_{+}$   & $-1$ & $-1$ & $1$              &     $\pi_0(C_{\delta-1})$  & $\mathbb{Z}$    & 0               & $\mathbb{Z}$    & 0               \\ 
$\mathrm{CI}_{+}$    & $1$  & $-1$ & $1$              &     $\pi_0(C_{\delta-1})$  & $\mathbb{Z}$    & 0               & $\mathbb{Z}$    & 0               \\ 
$\mathrm{CII}_{-}$   & $-1$ & $-1$ & $1$              &     $\pi_0(R_{\delta+2})$  & 0               & $\mathbb{Z}$    &  0              & 0               \\ 
$\mathrm{CI}_{-}$    & $1$  & $-1$ & $1$              &     $\pi_0(R_{\delta+6})$  & 0               & $\mathbb{Z}$    & $\mathbb{Z}_2$  & $\mathbb{Z}_2$  \\ \hline\hline
\end{tabular}
\end{center}
\caption{
Classification results for symmetry classes CII and CI. In these cases, systems preserve $CP'$ symmetry with $(CP')^2=1$ as well as $CP$-symmetry with $(CP)^2=-1$.
$(CP)(CP')=(CP')(CP)$ holds for symmetry classes $\mathrm{CII}_{+}$ and $\mathrm{CI}_{+}$, while $(CP)(CP')=-(CP')(CP)$ holds for symmetry classes $\mathrm{CII}_{-}$ and $\mathrm{CI}_{-}$.
}
\label{table: class_table_mech_CII_CI}
\end{table}

\subsubsection{
SPERs in mechanical graphene
}
In Sec.~\ref{sec: mech_fric}, we have seen that the mechanical graphene for $\Omega_0=0$ and $b_A=b_B$ hosts SPERs with extended chiral symmetry.
This fact can be seen from the classification results. The mechanical graphene preserves the $CP$-symmetry as well as the extended chiral symmetry. Thus, the symmetry class is BDI where the classification result is $\mathbb{Z}$ for $\delta=1$ (see Table~\ref{table: class_table_mech}).
Thus, the zero-th Chern number ($N_{0\mathrm{Ch}}\in \mathbb{Z}$) can be defined at each point of the BZ, indicating the presence of SPERs and SPESs in two- and three dimensions, respectively.
Therefore, the mechanical graphene with homogeneous friction exemplifies $\mathbb{Z}$-classification of class BDI with $\delta=1$.

Introducing inversion-symmetry-breaking terms changes the symmetry class from BDI to AIII. Table~\ref{table: class_table_mech} indicates that the SPERs in the mechanical graphene survive even in the absence of inversion symmetry; the classification result of class AIII for $\delta=1$ is $\mathbb{Z}$. 
This fact is consistent with our numerical simulation.

\subsubsection{
SPERs in rotating mechanical graphene
}
In Sec.~\ref{sec: rot_mech_fric} we have seen that the rotating mechanical system hosts SPERs with $CP$-symmetry when the frictional force is homogeneous.
This fact can be understood from the classification results. The rotating mechanical systems with inversion symmetry preserve $CP$-symmetry. Thus, the symmetry class is D where the classification result is $\mathbb{Z}_2$ for $\delta=1$ (see Table~\ref{table: class_table_mech}). 
Thus, the $\mathbb{Z}_2$-invariant~(\ref{eq: Z2inv}) can be defined at each point of the BZ, indicating the presence of SPERs and SPESs in two- and three dimensions.
Therefore, the rotating mechanical system with homogeneous friction exemplifies the $\mathbb{Z}_2$-classification of class D.
 
Introducing inhomogeneity breaks $CP$-symmetry and changes the symmetry class from D to A. Table~\ref{table: class_table_mech} shows that the classification results of class A is $0$ for $\delta=1$ and $\mathbb{Z}$ for $\delta=2$.
This indicates that the system may have a one-dimensional $\mathbb{Z}$-invariant taking a non-trivial value, although there is no zero-dimensional topological invariant. This result explains the instability of SPERs against inhomogeneity for the rotating mechanical graphene (see Sec.~\ref{sec: rot_mech_fric_w_inhomo}).

\subsubsection{
$\mathbb{Z}$-classification for $\mathrm{CII}_{+}$ and $\delta=1$
}
Tables~\ref{table: class_table_mech}~and~\ref{table: class_table_mech_CII_CI} indicate the following facts. 
For class $\mathrm{CII}_{+}$, there exist symmetry-protected non-Hermitian degeneracies characterized by a zero-dimensional $\mathbb{Z}$-invariant. 
These symmetry-protected non-Hermitian degeneracies are unstable against the perturbation breaking extended chiral symmetry (e.g., rotation) in contrast to the ones for class BDI.
This is because the perturbation changes the symmetry from class $\mathrm{CII}_{+}$ to class C where there is no topological invariant for $\delta=1$.

In the following, we discuss the above results by analyzing a $4\times4$ matrix.
Suppose that operators of $TP$- and $CP$-symmetry are given by
\begin{subequations}
\begin{eqnarray}
U_{TP} &=& i\rho_2\tau_0, \\
U_{CP}  &=& i\rho_2\tau_3, 
\end{eqnarray}
as well as these operators we assume that system preserves the $CP'$-symmetry whose operator is given by
\begin{eqnarray}
U_{CP'} &=&  \rho_0\tau_3.
\end{eqnarray}
\end{subequations}
A $4\times 4$ Hamiltonian satisfying the above symmetry is written as
\begin{eqnarray}
\label{eq: 4x4_Hami_CII+}
 H_{\mathrm{CII}_+} &=& id_{00}\rho_0\tau_0+b_{01}\rho_0\tau_1+b_{22}\rho_2\tau_2+id_{03}\rho_0\tau_3.
\end{eqnarray}
For $b_{00}=b_{22}=0$, the above Hamiltonian describes the one-dimensional subsystem ($k_x=k_y$) of a mechanical square lattice (see Appendix~\ref{sec: deri_sq_mech}).
The eigenvalues of the Hamiltonian can be obtained as
\begin{eqnarray}
 E &=& id_{00} \pm \sqrt{b^2_{01}+b^2_{02}-d^2_{03}},
\end{eqnarray}
by noticing that it commutes with $U:=U_{CP}U_{CP'}=\rho_2\tau_0$.
Therefore, satisfying only one condition results in the defective Hamiltonian (i.e., the non-diagonalizable Hamiltonian). This fact indicates the presence of exceptional points in the one-dimensional subsystem ($k_x=k_y$) of the mechanical square lattice, which corresponds to the case with $(d,d_{EP})=(1,0)$.

Now, we add an perturbation breaking both of $TP$- and extended chiral symmetry,
\begin{eqnarray}
\label{eq: ptb_Hc}
 H'&=& b_{20}\rho_2\tau_0 +b_{23}\rho_2\tau_3+id_{02}\tau_2 \rho_0 + id_{21}\rho_2\tau_1.
\end{eqnarray}
The symmetry class of the resulting Hamiltonian ($H_{\mathrm{C}}:=H_{\mathrm{CII}_{+}}+H'$) is class C. 
Because the system preserves $CP$- and $CP'$-symmetry, we can block-diagonalize the Hamiltonian $H_{\mathrm{C}}$ with the matrix $U$~\cite{footnote_blockdiag_CII+}.
Thus, the resulting eigenvalues of this model are written as
\begin{eqnarray}
 E_{s} &=& (sb_{20}+id_{00}) \pm \sqrt{b'^2-d'^2+2is\bm{b}'\cdot\bm{d}'},
\end{eqnarray}
with $\bm{b}'=(b_{01},b_{22},b_{23})$ and $\bm{b}'=(d_{21},d_{02},d_{03})$. Here, $s$ takes $\pm 1$, labeling each sector of $U$.
Therefore, satisfying both of two conditions, $b'^2-d'^2=0$ and  $\bm{b}'\cdot\bm{d}'=0$, result in the defective Hamiltonian.
This fact indicates that exceptional points in the subsystem ($k_x=k_y$) of the mechanical square lattice are unstable against rotation because rotating the system introduces the Coriolis force described by $b_{23}\rho_2\tau_3$ in Eq.~(\ref{eq: ptb_Hc}).

\section{
Summary
}
We have elucidated that mechanical systems host symmetry-protected non-Hermitian degeneracies (e.g., SPERs) due to the emergent symmetry~(\ref{eq: chiral}). This fact indicates that mechanical systems are suited experimental platforms of symmetry-protected non-Hermitian degeneracies because fine-tuning is not required in contrast to other experimental platforms (e.g., photonic systems).
Specifically, we have demonstrated the emergence of the SPERs for the mechanical graphene with homogeneous friction. In this system, the zero-th Chern number, characterizing SPERs, can be obtained only from experimentally observable quantities; the coefficient of the fiction and the frequency for the case without dissipation. 
In addition, inside of the SPERs, there exists a mode damping slowly compared to the one outside of the ring, which can be an experimental signal of SPERs.

Furthermore, we have carried out topological classification by taking into account the special characteristics of mechanical systems in order to systematically understand the interplay of non-Hermitian degeneracies and the emergent symmetry. The obtained results elucidate the presence of SPERs with $CP$-symmetry in rotating mechanical graphene. The obtained results also predict symmetry-protected non-Hermitian degeneracies in two- and three dimensions whose demonstration for specific mechanical systems is left as a future work.

\section{
Acknowledgement
}
This work is partly supported by JSPS KAKENHI Grants 
No.~JP16K13845, 
No.~JP17H06138, 
and No.~JP18H05842. 
A part of numerical calculations were performed on the supercomputer at the ISSP in the University of Tokyo.


\begin{thebibliography}{54}
\bibitem [{foo({\natexlab{a}})}]{footnote_2x2chiral}%
  \BibitemOpen
  \href@noop {} {}\bibinfo {note} {{ 
  In order to satisfy $U^2_{\Gamma}=\sigma_0$, the projected operator
  $U_{\Gamma}$ takes $U_{\Gamma}=\bm{n}\cdot \bm{\sigma}$ or $\sigma_0$.
  However when $U_{\Gamma}=\sigma_0$, the projected Hamiltonian becomes
  anti-Hermitian whose eigenvalues are always pure imaginary }}\BibitemShut
  {NoStop}%
\bibitem [{foo({\natexlab{b}})}]{footnote_ERchiral_b=d=0}%
  \BibitemOpen
  \href@noop {} {}\bibinfo {note} {{ 
  We note that when $b=d=0$ holds, the Hamiltonian is not defective. However,
  this condition is not satisfied without fine-tuning }}\BibitemShut {NoStop}%
\bibitem [{foo({\natexlab{c}})}]{footnote_bandtouch}%
  \BibitemOpen
  \href@noop {} {}\bibinfo {note} {{ 
  The band touching occur when the two band touches in the complex plane
  because the Hamiltonian is non-Hermitian }}\BibitemShut {NoStop}%
\bibitem [{foo({\natexlab{d}})}]{footnote_blockdiag_CII+}%
  \BibitemOpen
  \href@noop {} {}\bibinfo {note} {{ 
  The block-diagonalized Hamiltonian for the $\pm$-sector is written as
  \begin{eqnarray} H_{\pm} &=& b_{00}\tau_0+b_{01}\tau_1\pm
  b_{22}\tau_2+id_{03}\tau_3\nonumber \end{eqnarray} }}\BibitemShut {NoStop}%
\end{thebibliography}
%

\appendix
\section{
Symmetry of the matrix $Q(\bm{k})$
}
\label{sec: symm Q(k)}

We show that the matrix $Q(\bm{k})$, satisfying $D(\bm{k})=Q^2(\bm{k})$, inherits symmetry of $D(\bm{k})$.
Firstly, we note that the matrix $Q$ can be specifically obtained by diagonalizing $D$;
\begin{subequations}
\label{eq: spefific_Q_app}
\begin{eqnarray}
  Q(\bm{k})&=&U(\bm{k}) \Lambda(\bm{k})^{1/2} U^\dagger(\bm{k}),  \\
  D(\bm{k})&=& U(\bm{k}) \Lambda(\bm{k}) U^\dagger(\bm{k}),
\end{eqnarray}
\end{subequations}
where $U(\bm{k})$ is a unitary matrix. $\Lambda(\bm{k})$ is a diagonal matrix whose elements are eigenvalues of the positive semidefinite matrix $D(\bm{k})$.

Now, we show that the following two relations hold.

\begin{subequations}
\label{eq: symm_D_to_Q_inv}
For 
\begin{eqnarray}
U_ID(\bm{k})U_I &=& D(-\bm{k}),
\end{eqnarray}
the matrix $Q$ satisfies
\begin{eqnarray}
U_IQ(\bm{k})U_I &=& Q(-\bm{k}).
\end{eqnarray}
\end{subequations}
%
\begin{subequations}
\label{eq: symm_D_to_Q_PH}
For 
\begin{eqnarray}
D^*(\bm{k}) &=& D(-\bm{k}),
\end{eqnarray}
the matrix $Q$ satisfies
\begin{eqnarray}
Q^*(\bm{k}) &=& Q(-\bm{k}).
\end{eqnarray}
\end{subequations}

Firstly, we prove Eq.~(\ref{eq: symm_D_to_Q_inv}b). As the condition~(\ref{eq: symm_D_to_Q_inv}a) hold, we have
\begin{eqnarray}
 U_I Q(\bm{k}) U_I &=& U_I [D^{1/2}(\bm{k})]U_I \nonumber \\
                                        &=&  [U_ID(\bm{k})U_I]^{1/2} \nonumber \\
                                        &=&  [D(-\bm{k})]^{1/2} \nonumber \\
                                        &=&  Q(-\bm{k}),
\end{eqnarray}
which results in Eq.~(\ref{eq: symm_D_to_Q_inv}b).

In a similar way, with Eq.~(\ref{eq: symm_D_to_Q_PH}a), we have 
\begin{eqnarray}
  Q^*(\bm{k})&=& [D^{1/2}(\bm{k})]^* \nonumber \\
                    &=& [D^*(\bm{k})]^{1/2}  \nonumber \\
                    &=& [D(-\bm{k})]^{1/2}  \nonumber \\
                    &=& Q^*(\bm{k}),
\end{eqnarray}
which results in Eq.~(\ref{eq: symm_D_to_Q_PH}b).

\section{
Eigenvalues of Hamiltonian with extended chiral symmetry
}
\label{sec: eivenval_chiral_app}
Here, we show that in the presence of the extended chiral symmetry ($U_\Gamma H^\dagger U^\dagger_\Gamma=-H$), the energy eigenvalues form a pair $(E,-E^*)$ or take pure imaginary values. 

Suppose that $| \phi^n_R \rangle$ and $| \phi^n_L \rangle$ are right and left eigenvectors of $H$;
\begin{subequations}
\label{eq: def_left_right_eiven}
\begin{eqnarray}
 H | \phi^n_R \rangle &=& E_n | \phi^n_R \rangle, \\
 H^\dagger | \phi^n_L \rangle &=& E^*_n | \phi^n_L \rangle.
\end{eqnarray}
\end{subequations}
Then, we have 
\begin{eqnarray}
 H^\dagger U_\Gamma | \phi^n_R \rangle = - U_\Gamma  H | \phi^n_R \rangle = - E_n U_\Gamma | \phi^n_R \rangle.
\end{eqnarray}
Comparing with Eq.~(\ref{eq: def_left_right_eiven}b), we see that the energy eigenvalues form a pair $(E,-E^*)$ or are pure imaginary.

\section{
Symmetry-protected non-Hermitian degeneracies with $CP$-symmetry
}
\label{sec: protection cp}
\subsection{
Analysis of a $2\times2$ matrix around the band touching point
}

Now, we elucidate that $CP$-symmetry with $CP^2=1$ protects the SPERs in the two-dimensional BZ. 
(The following argument also holds for three-dimensional systems which host SPESs.)

Firstly, we note that in the presence of $CP$-symmetry~(\ref{eq: CP}), two eigenvalues of the Hamiltonian form a pair $(E,-E^*)$ or become pure imaginary. 
This can be seen as follows. Consider the right eigenstate $|\phi^n_R\rangle$ of the Hamiltonian $H$ with the eigenvalue $E$. Then, we obtain
\begin{eqnarray}
 H U|\phi^n_R\rangle^* =-U H^* |\phi^n_R\rangle^* = - E^* U|\phi^n_R\rangle^*,
\end{eqnarray}
where we have used Eq.~(\ref{eq: CP}).
The above fact indicates that each pair $(E,-E^*)$ cannot split without going through a band touching.

Now, let us consider the case where two energy bands touch at a point in the BZ.
Around the point the Hamiltonian can be represented as a $2\times 2$ matrix which is generically given by
\begin{eqnarray}
H_{2\times2}(\bm{k}) &=& \sum_{\alpha=0,\cdots,3} (b_\alpha+id_\alpha) \sigma_\alpha.
\end{eqnarray}
$\sigma$'s are the Pauli matrices which act on the two states showing the band touching.
In this two-dimensional Hilbert space, the $CP$ operator is represented as $CP=\mathcal{K}$ ($CP=\sigma_2 \mathcal{K}$) for $CP^2=1$ ($CP^2=-1$), respectively. The derivation is shown in Appendix~\ref{sec: rep_CP_app}.
In the presence of $CP$-symmetry with $CP=\mathcal{K}$, the generic Hamiltonian is written as
\begin{eqnarray}
H_{2\times2}(\bm{k}) &=&
\left(
\begin{array}{cc}
i(d_0+d_3) & id_1-ib_2 \\
id_1+ib_2 & i(d_0-d_3)
\end{array}
\right).
\end{eqnarray}
Diagonalizing the matrix yields
\begin{subequations}
\begin{eqnarray}
E_{\pm}(\bm{k}) &=& id_0 \pm \sqrt{ b^2_2 -(d^2_1+d^2_3)},
\end{eqnarray}
indicating that the condition of the band touching is 
\begin{eqnarray}
\label{eq: cond_BT_CP_plus1}
b^2_2=d^2_1+d^2_3.
\end{eqnarray}
\end{subequations}
At the band touching point, the Hamiltonian describing these two bands becomes defective because the Hamiltonian can be written as 
\begin{eqnarray}
H_{2\times2}(\bm{k}) &=& 
|b_2|
\left(
\begin{array}{cc}
0 &  1\\
0 &  0
\end{array}
\right).
\end{eqnarray}
For $b^2=d^2_1+d^2_3=0$, the Hamiltonian becomes Hermitian and is diagonalizable. However, that such condition cannot be satisfied without fine-tuning. 

We note that the condition of the band touching~(\ref{eq: cond_BT_CP_plus1}) specifies a one-dimensional line in the BZ which corresponds to SPERs.

For $CP=\sigma_2\mathcal{K}$, the eigenvalue problem is reduced to the one for a  Hermitian matrix because the $CP$-symmetry results in
\begin{eqnarray}
H_{2\times2}(\bm{k}) &=&id_0\sigma_0+ \sum_{\mu=1,2,3} b_\mu \sigma_\mu.
\end{eqnarray}

\subsection{
Representation of the $CP$-transformation
}
\label{sec: rep_CP_app}
We show that the operator of $CP$-symmetry is represented as 
\begin{eqnarray}
CP &=& \mathcal{K},
\end{eqnarray}
for $CP^2=1$ and 
\begin{eqnarray}
CP &=& \sigma_2 \mathcal{K}, 
\end{eqnarray}
for $CP^2=-1$.

This can be seen as follows.
A generic unitary matrix $U_{CP}$ can be represented as 
\begin{eqnarray}
U_{CP} &=& e^{i\bm{b}\cdot\bm{\sigma}},
\end{eqnarray}
with $\bm{b}\in \mathbb{R}^3$ up to the global phase factor.
Thus, we can see
\begin{eqnarray}
U_{CP}U^*_{CP}  &=& [\cos^2(b)+\sin^2(b) \hat{\bm{b}}\cdot\hat{\bm{b}}'] \nonumber \\
                &&+i[\sin^2b (\bm{b}\times\bm{b}') +\frac{1}{2}\sin(2b)(\hat{\bm{b}}-\hat{\bm{b}}')  ] \cdot \bm{\sigma}, \nonumber \\
\end{eqnarray}
with $\bm{b}:=(b_1,b_2,b_3)$ and $\bm{b}':=(b_1,-b_2,b_3)$. $\hat{\bm{b}}$ ($\hat{\bm{b}}'$) is a unit vector proportional to $\hat{\bm{b}}$ ($\hat{\bm{b}}'$) respectively.

Because $U_{CP}U^*_{CP}$ is proportional to the identity operator, we see that one of the following conditions are satisfied:
(i) $b = n\pi$ with $n=0,1,2,\cdots$, (ii) $\bm{b} = (b_1,0,b_3)^T$, and (iii) $\bm{b} = (0,b_2,0)^T$.

For case (i) and (ii), we have $U_{CP}=\sigma_0$.
For case (iii), we have $U_{CP}=\sigma_2$.

For case (ii), we note that $\bm{b}\cdot\bm{\sigma}$ with $\bm{b} = (b_1,0,b_3)^T$ is a real-symmetric matrix which can be diagonalized with an orthogonal matrix.
Thus, without loss of generality, we have $U_{CP}=\sigma_0=\cos(b)\sigma_0+i\sin(b)\sigma_3$ which is reduced to $U_{CP}=\sigma_0$ by a proper choice of the gauge.

\section{
Details of the mechanical graphene
}
\label{sec: deri_mech_grah}
Here, we derive Eq.~(\ref{eq: L of mech_graph}). 
We suppose that the dissipation and the internal force are absent and that the spring-mass forms the honeycomb lattice (Fig.~\ref{fig: scketch of lattice}).
\begin{figure}[!h]
\begin{center}
\includegraphics[width=90mm,clip]{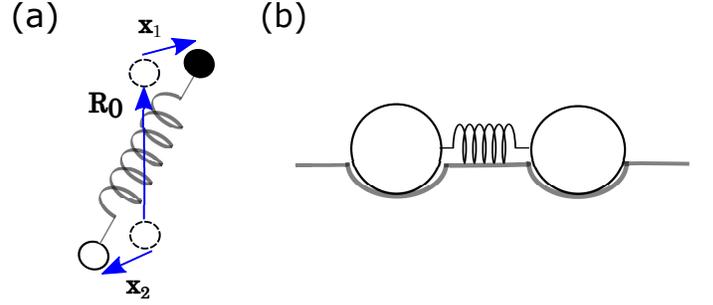}
\end{center}
\caption{(Color Online).
Sketch of the system.
(a) Two mass points connected with a spring. (b) Mass points trapped by dents of the floor.
}
\label{fig: sckech_model_app}
\end{figure}
As a first step, we focus on the two mass points of this spring-mass system [Fig.~\ref{fig: sckech_model_app}(a)]. 
In this case, the potential energy arises from the spring is described as 
\begin{eqnarray}
U_s &=& \frac{\kappa}{2} (\sqrt{R^2_0+\delta \bm{x}}-l_0),
\end{eqnarray}
where $l_0$ denotes natural length of the spring, and $\kappa$ denotes a spring constant. 
Two dimensional vectors are shown in Fig~\ref{fig: sckech_model_app}(a); $\bm{R}_0$ connects two lattice points; $\delta \bm{x}:=\bm{x}_1-\bm{x}_2$ where $\bm{x}_i$ $(i=1,2)$ describes the displacement of mass points.
Supposing that $\delta x \ll R_0$, the potential energy can be expanded as
\begin{eqnarray}
U_s &=& \frac{\kappa}{2} \left[(R_0-l_0)^2 \right. \nonumber \\
    && \left. +2(1-\eta^{-1}) \bm{R}\cdot \delta \bm{x}+\delta x_\mu \gamma^{\mu\nu}_{\bm{R}_0} \delta x_\nu \right], 
\end{eqnarray}
with $\mu=x,y$, $\hat{\bm{R}}_0=\bm{R}_0/R_0$, and $\gamma^{\mu\nu}_{\bm{R}_0}=(1-\eta)\delta^{\mu\nu}+\eta \hat{R}^\mu_0\hat{R}^\nu_0$. 

Therefore, the Lagrangian of the mass points with mass $m$ is given by
\begin{subequations}
\label{eq: L_app}
\begin{eqnarray}
 \mathcal{L}&=& T-U_1-U_2, \\
 T &=& \frac{m}{2}\sum_{i} \dot{\bm{x}}^2_i,\\
 U_1 &=&  \sum_{i} \frac{\kappa^0_i}{2} \bm{x}^2_i,\\
 U_2 &=& \frac{\kappa}{2}\sum_{\langle i j \rangle} \left[(R_{ij}-l_0)^2+2(1-\eta^{-1}) \bm{R}_{ij}\cdot (\bm{x}_i-\bm{x}_j) \right. \nonumber \\
    && \quad+(x_{i\mu}-x_{j\mu}) \gamma^{\mu\nu}_{\bm{R}_{ij}} (x_{i\nu}-x_{j\nu}) ].
\end{eqnarray}
\end{subequations}
Here, we assume that the mass points are trapped on dents of the floor [Fig.~\ref{fig: sckech_model_app}(b)], which induces the potential term $U_1$.
$\kappa^0_i$ describes strength of the gravitational energy arising from the dents of the floor. $\kappa^0_i=\kappa^0_\alpha$ when the site $i$ belongs to sublattice $\alpha$.
$U_2$ describes the potential energy arising from the springs.
$i$ and $j$ label mass points.  $R_{i}$ denotes the equilibrium position of mass point $i$. $R_{ij}:=\bm{R}_i-\bm{R}_j$.
The second term of Eq.~(\ref{eq: L_app}d) is zero as long as the equilibrium is achieved for $\bm{x}_{i}=0$. Thus, we omit it.

Applying Fourier transformation,
\begin{eqnarray}
 \bm{u}_{\bm{k}\alpha}(t)&=& \frac{1}{\sqrt{N}} \sum_{i \in \alpha } e^{i \bm{k} \cdot \bm{\tilde{R}}_i} \bm{x}_{\tilde{R}_i\alpha}(t),
\end{eqnarray}
the Lagrangian is rewritten as
\begin{subequations}
\label{eq: Lu_app}
\begin{eqnarray}
\mathcal{L}
 &=&
\frac{m}{2}
\sum_{\bm{k},\mu,\alpha}  \dot{u}^{\mu}_{\bm{k}\alpha} \cdot \dot{u}^{\mu}_{-\bm{k}\alpha} - \frac{1}{2}\sum_{\bm{k}}  D^{\mu\nu}_{\alpha\beta}(\bm{k}) u^\mu_{\bm{k},\alpha} u^\nu_{-\bm{k},\beta}, \nonumber\\
\end{eqnarray}
with $D^{\mu\nu}_{\alpha\beta}(\bm{k})=D_{\mu\alpha;\nu\beta}(\bm{k})$ and
\begin{eqnarray}
D_{\mu\alpha;\nu\beta}(\bm{k})
 &=& 
 3\kappa(1-\frac{\eta}{2})\1+
 \mat{  \kappa^0_A & D_{AB}(\bm{k})  \\  D^\dagger_{AB} (\bm{k})  & \kappa^0_B},
\nonumber 
\\
&& \\
D^{\mu\nu}_{AB}(\bm{k}) &=& 
-\kappa( \gamma_3+ \gamma_1 e^{-i\bm{k}\cdot \bm{a}_1}+ \gamma_2 e^{-i\bm{k}\cdot \bm{a}_2 }),
\end{eqnarray}
\begin{eqnarray}
\gamma_1&:=& (1-\eta) \mat{ 1 & 0 \\  0 & 1} + \eta \mat{ \frac{3}{4} & \frac{\sqrt{3}}{4}  \\  \frac{\sqrt{3}}{4}  & \frac{1}{4} } , \\
\gamma_2&:=& (1-\eta) \mat{ 1 & 0 \\  0 & 1} + \eta \mat{ \frac{3}{4} & -\frac{\sqrt{3}}{4}  \\  -\frac{\sqrt{3}}{4}  & \frac{1}{4} } , \\
\gamma_3&:=& (1-\eta) \mat{ 1 & 0 \\  0 & 1} + \eta \mat{ 0 & 0  \\  0  & 1 } . 
\end{eqnarray}
\end{subequations}
Here, $\bm{x}_{\tilde{R}_i\alpha}$ denotes the displacement of the mass point at $\alpha(=A,B)$-sublattice of the unit cell specified by $\bm{\tilde{R}}_i$.
The above results correspond to Eq.~(\ref{eq: L of mech_graph}).

\section{
Simplified zero-th Chern number for mechanical systems
}
\label{sec: 0Ch_app}
Firstly, we define the zero-th Chern number. Consider a traceless non-Hermitian Hamiltonian satisfying the extended chiral symmetry,
\begin{eqnarray}
 U_\Gamma H^\dagger U^\dagger_\Gamma &=&-H,
\end{eqnarray}
where $U_\Gamma$ is an unitary matrix satisfying $U^2_\Gamma=\1$.
In this case, the zero-th Chern number counts the number of the occupied bands ($\epsilon_n<0$) where $\epsilon_n$ with $n=1,\cdots,\mathrm{dim}H$ denotes the eigenvalue of the Hermitian matrix $-iHU_\Gamma$.

The simplified form of the zero-th Chern number~(\ref{eq: simp_0Ch_mech}) can be obtained as follows.
In the similar way as Sec.~\ref{sec: mapping EOM}, for $\Gamma_0=-b\1$ with $b>0$, we can obtain the traceless Hamiltonian describing the mechanical system
\begin{eqnarray}
 H &=&
\left(
\begin{array}{cc}
i\frac{b}{2}\1 &  Q(\bm{\bm{k}})\\
Q(\bm{\bm{k}}) & -i\frac{b}{2}\1
\end{array}
\right),
\end{eqnarray}
preserving the extended chiral symmetry with $\Gamma=\rho_3$.
Here, the matrix $Q$ is defined as $D=Q^2$ with the matrix $D$ describing the potential force [see Eq.~(\ref{eq: spefific_Q_app})].
Namely, the eigenvalue of the matrix $Q$ corresponds to the frequency of the oscillation $\omega_{0n}(\bm{k})$.
Diagonalizing the matrix $Q$, the matrix $-iH\rho_3$ is rewritten as
\begin{eqnarray}
-iH\rho_3 &=& \bigoplus_n 
\left(
\begin{array}{cc}
\frac{b}{2} & i\omega_{0n} \\
-i\omega_{0n} & \frac{b}{2}
\end{array}
\right).
\end{eqnarray}
Correspondingly, the eigenvalues $\epsilon_n$ is written as
\begin{eqnarray}
 \epsilon_n(\bm{k}) &=& \frac{b}{2}\pm \omega_{0n}(\bm{k}).
\end{eqnarray}
Therefore, the zero-th Chen number is obtained as
\begin{eqnarray}
 N_{0\mathrm{Ch}}(\bm{k})&=& \sum_{n=1,\cdots,\mathrm{dim}H} \Theta_>(2\omega_{0n}(\bm{k})-b),
\end{eqnarray}
where $\Theta_>(x)$ takes $1$, $1/2$, and $0$ for $x>0$, $x=0$, and $x<0$, respectively.
Notably, $N_{0\mathrm{Ch}}$ is obtained with the experimental observables for the mechanical system, $\omega_{0n}$ and $b$.

\section{
Mechanical square lattice
}
\label{sec: deri_sq_mech}
In a similar way to Appendix.~\ref{sec: deri_mech_grah}, we can obtain the Lagrangian of a spring-mass model forming a square lattice.
The matrix $D(\bm{k})$ describing the potential force is written as
\begin{eqnarray}
D(\bm{k}) &=& 
4\kappa(1-\frac{\eta}{2})\rho_0
-2\kappa \left[(1-\eta)( \cos(k_x)+\cos(k_y) )\rho_0 \right.
\nonumber\\
&&
\left.
+\eta\cos(k_x) 
\left(
\begin{array}{cc}
1 & 0 \\
0 & 0
\end{array}
\right)_\rho
+\eta\cos(k_y) 
\left(
\begin{array}{cc}
0 & 0 \\
0 & 1
\end{array}
\right)_\rho
\right]\nonumber \\
&=&
2\kappa(1-\frac{\eta}{2})[2-2\cos(\frac{k_x+k_y}{2})\cos(\frac{k_x-k_y}{2})]\rho_0 \nonumber \\
&&+2\kappa\eta\sin(\frac{k_x+k_y}{2})\sin(\frac{k_x-k_y}{2})\rho_3.
\end{eqnarray}
For $k_x-k_y=0$, we can see that the term proportional to $\rho_3$ vanishes.
Therefore, for system with a homogeneous frictional force, rewriting the equation of motion with Hamiltonian [see Eq.~(\ref{eq: BdG gen})], we can see that the obtained Hamiltonian corresponds to Eq.~(\ref{eq: 4x4_Hami_CII+}) 
with
\begin{eqnarray}
b_{00} &=& 0, \\ 
b_{01} &=& \sqrt{ 4\kappa(1-\frac{\eta}{2})(1-\cos(k_x))},\\
b_{22} &=& 0, 
\end{eqnarray}
where $b_{03}$ in Eq.~(\ref{eq: 4x4_Hami_CII+}) describes the strength of the frictional force.
Here, we have set the trap potential illustrated in Fig.~\ref{fig: sckech_model_app}(b) is zero.

\section{
Another type of $CP$-symmetry
}
\label{sec: CP_transp_app}
Non-Hermiticity of the Hamiltonian results in two types of particle-hole symmetry~\cite{Kawabata_gapped_class_arXiv19}. 
One is given by Eq.~(\ref{eq: PH}) and the other is 
\begin{eqnarray}
\label{eq: PHd_app}
\rho_3 H^T(\bm{k}) \rho_3 &=& -H(-\bm{k}).
\end{eqnarray}
However, we note that $CP$-symmetry, defined as the product of particle-hole symmetry~(\ref{eq: PHd_app}) and inversion symmetry, is incompatible to mechanical systems with homogeneous friction which is one of the simplest setups for non-Hermitian mechanical systems.
When the Hamiltonian preserves the $CP$-symmetry,
\begin{eqnarray}
\label{eq: CPd}
U_{CP} H^T(\bm{k}) U^\dagger_{CP} &=& -H(\bm{k}),
\end{eqnarray}
with $U_{CP}=U_I\otimes \rho_3$ and $U^2_I=\1$, we obtain
\begin{subequations}
\begin{eqnarray}
U_IQ^T(\bm{k})U_I &=& Q(\bm{k}), \\
U_I\Gamma^T_0(\bm{k})U_I &=& -\Gamma_0(\bm{k}), 
\end{eqnarray}
\end{subequations}
For homogeneous friction $\Gamma_0=-b\1$, the second equation yields $b=0$,
indicating that $CP$-symmetry~(\ref{eq: CPd}) is incompatible to mechanical systems with homogeneous friction.

\section{
Ten-fold way classification of Hermitian Hamiltonian for AZ+$\mathcal{I}$ classes with additional chiral symmetry
}
\label{sec: class_Hermi_Sig_app}
We address the topological classification of Hermitian topological insulators/superconductors in the presence or absence of $TP$-, $CP$-, and chiral symmetry by taking into account additional chiral symmetry. 
The classification without additional chiral symmetry has been addressed in Ref.~\onlinecite{Bzdusek_AZ+I_PRB17} where the symmetry classes are called AZ+$\mathcal{I}$ symmetry classes.

Here, we suppose $d$-dimensional systems to satisfy
\begin{subequations}
\begin{eqnarray}
\{ H(\bm{k}), \Sigma \} &=&0,
\end{eqnarray}
as well as constraints of each symmetry class. Here, the Hamiltonian preserving $TP$-, $CP$-, and chiral symmetry satisfies
\begin{eqnarray}
(TP) H(\bm{\bm{k}}) (TP)^{-1} &=& H(\bm{\bm{k}}), \\
(CP) H(\bm{\bm{k}}) (CP)^{-1} &=& -H(\bm{\bm{k}}), \\
\Gamma H(\bm{\bm{k}}) \Gamma^{-1} &=& -H(\bm{\bm{k}}),
\end{eqnarray}
where the operators of $TP$- and $CP$-transformations are anti-unitary while the operator of the chiral transformation is unitary.
\end{subequations}
We note that the operators $TP$, $CP$, and $\Gamma$ defined above correspond to $\widetilde{TP}$, $\widetilde{CP}$, and $\widetilde{\Gamma}$ defined in Eq.~(\ref{eq: all symm tilde}).
(In order to simplify the notation, we have omitted the tilde.)

In this section, we assume that the additional chiral symmetry satisfies the following relations
\begin{subequations}
\label{eq: comm_TP_CP_Sigma}
\begin{eqnarray}
\{TP,\Sigma\} &=& 0, \\
{}[CP,\Sigma] &=& 0, \\
\{\Gamma,\Sigma\} &=& 0,
\end{eqnarray}
\end{subequations}
which hold for the generic mechanical systems with friction (see Sec.~\ref{sec: class_scheme_nonH}).

The classification results are summarized in Table~\ref{table: Hermi_class_AZ+I}.
%
%
%
\begin{table}[htb]
\begin{center}
\begin{tabular}{c c c c c c c c c c c c c} \hline\hline
     & $TP$ & $CP$ & $\Gamma$ & homotopy                       & $d=0$           & 1              &    2           & 3              & 4              & 5              & 6              &    7            \\ \hline
A    & $0$  & $0$  & $0$              &     $\pi_0(C_{d+1})$           &    0            & $\mathbb{Z}$   &    0           & $\mathbb{Z}$   & 0              & $\mathbb{Z}$   & 0              & $\mathbb{Z}$    \\ 
AIII & $0$  & $0$  & $1$              &     $\pi_0(C_{d})$             & $\mathbb{Z}$    &    0           & $\mathbb{Z}$   & 0              & $\mathbb{Z}$   & 0              & $\mathbb{Z}$   & 0               \\ \hline
AI   & $1$  & $0$  & $0$              &     $\pi_0(R_{d+7})$           &    0            & $\mathbb{Z}$   & $\mathbb{Z}_2$ & $\mathbb{Z}_2$ & 0              & $\mathbb{Z}$   &    0           &    0            \\ 
BDI  & $1$  & $1$  & $1$              &     $\pi_0(R_{d})$             &  $\mathbb{Z}$   & $\mathbb{Z}_2$ & $\mathbb{Z}_2$ &    0           & $\mathbb{Z}$   &    0           &    0           &    0            \\ 
D    & $0$  & $1$  & $0$              &     $\pi_0(R_{d+1})$           &  $\mathbb{Z}_2$ & $\mathbb{Z}_2$ &    0           & $\mathbb{Z}$   &    0           &    0           &    0           &  $\mathbb{Z}$   \\ 
DIII & $-1$ & $1$  & $1$              &     $\pi_0(R_{d+2})$           &  $\mathbb{Z}_2$ &    0           & $\mathbb{Z}$   &    0           &    0           &    0           &  $\mathbb{Z}$  &  $\mathbb{Z}_2$ \\ 
AII  & $-1$ & $0$  & $0$              &     $\pi_0(R_{d+3})$           &    0            & $\mathbb{Z}$   &    0           &    0           &    0           &  $\mathbb{Z}$  &  $\mathbb{Z}_2$&  $\mathbb{Z}_2$ \\ 
CII  & $-1$ & $-1$ & $1$              &     $\pi_0(R_{d+4})$           &  $\mathbb{Z}$   &    0           &    0           &    0           &  $\mathbb{Z}$  &  $\mathbb{Z}_2$&  $\mathbb{Z}_2$&    0            \\ 
C    & $0$  & $-1$ & $0$              &     $\pi_0(R_{d+5})$           &    0            &    0           &    0           &  $\mathbb{Z}$  &  $\mathbb{Z}_2$&  $\mathbb{Z}_2$&    0           &  $\mathbb{Z}$   \\
CI   & $1$  & $-1$ & $1$              &     $\pi_0(R_{d+6})$           &    0            &    0           & $\mathbb{Z}$   & $\mathbb{Z}_2$ &  $\mathbb{Z}_2$&   0            &  $\mathbb{Z}$  &    0            \\ \hline \hline
\end{tabular}
\end{center}
\caption{
Classification results for Hermitian systems of AZ+$\mathcal{I}$ symmetry classes with additional chiral symmetry.
``$0$" in the second, the third, and the fourth columns denotes that the corresponding symmetry is absent. The $\pm 1$ in the second [the third] column represents the sign of $(TP)^2=\pm 1$ [$(CP)^2=\pm 1$], respectively. 
From sixth to thirteenth columns, the classification results are summarized where $\mathbb{Z}$ or $\mathbb{Z}_2$ means the presence of topological phases with the corresponding topological invariant. ``0” in these columns indicates the absence of topological phases.
}
\label{table: Hermi_class_AZ+I}
\end{table}
%
%
%
In this table, we can see that the classification results for AZ$+\mathcal{I}$ symmetry classes with additional chiral symmetry are given by $\pi(C_{q+d})$ and $\pi(R_{q+d})$ while those for ordinary AZ symmetry classes are given by $\pi(C_{q-d})$ and $\pi(R_{q-d})$.
This difference arises from the fact that the inversion flips the momentum; $\bm{k}\to-\bm{k}$.

In the following, we see the details of the calculations.

\subsection{
Classification scheme
}
The topological classification of the Hermitian Hamiltonian is accomplished by the following steps.

\begin{enumerate}
 \item 
Deform the Hamiltonian into the gapped Dirac Hamiltonian
\begin{eqnarray}
H &=& \sum_{j=1,\cdots,d} k_j \gamma_j +m\gamma_0.
\end{eqnarray}
Here, $ \{ \gamma_i,\gamma_j \}=2\delta_{i,j}$ holds for $i,j=0,\cdots,d$.
When the Hamiltonian can be block-diagonalized with an unitary operator, we consider the above Dirac Hamiltonian for each sector.
\item 
 Consider a Clifford algebra $Cl_q$ or $Cl_{p,q}$ with symmetry operators and the matrices describing kinetic terms.
 Here, $Cl_q$ denotes a complex Clifford algebra having $q$ generators,
\begin{eqnarray}
\{e_1,e_2,\cdots,e_q\},
\end{eqnarray}
with $e^2_i=1$ for $i=1,\cdots,q$.
$Cl_{p,q}$ denotes a real Clifford algebra having $p+q$ generators,
\begin{eqnarray}
\{e_1,\cdots,e_p;e_{p+1},\cdots,e_{p+q}\},
\end{eqnarray}
where $p$ ($q$) generators square to $-1$ ($1$); $e^2_i=1$ ($-1$) for $i=1,\cdots,p$ ($i=p+1,\cdots,p+q$), respectively.

Because $TP$- and $CP$-transformations are described by anti-unitary operators, we need to introduce a operator $J$ to describe the complex structure in the presence of $TP$- or $CP$-symmetry (i.e., complex structure of real classes).
Here, the operator $J$ satisfies the relations; $J^2=-1$, $\{ TP, J\}=\{CP, J\}=[H(\bm{k}),J]=0$.
 \item 
 Consider the extension problem by adding the mass term $\gamma_0$ in order to obtain the corresponding classifying space.
 When the extension problem is written as $Cl_q \to Cl_{q+1}$ [$Cl_{p,q}\to Cl_{p,q+1}$], the corresponding classifying space is $C_q$ [$R_{q-p}$], respectively. 

 We note that when the extension problem is written as $Cl_{p,q}\to Cl_{p+1,q}$, the corresponding classifying space is $R_{2+p-q}$.
\item Obtain the classification result $\pi_0(C_q)$ [$\pi_0(R_q)$]. Here we use the fact listed in Table~\ref{table: homotopy}.
  We note that the Bott periodicity holds; $\pi_0(C_q)=\pi_0(C_{q+2})$ and $\pi_0(R_q)=\pi_0(R_{q+8})$.
\end{enumerate}

\begin{table}[htb]
\begin{center}
\begin{tabular}{c c} \hline\hline
classifying space & $\pi_0(C_q)$ or $\pi_0(R_q)$  \\ \hline
$C_0$ & $\mathbb{Z}$   \\
$C_1$ & $0$            \\ \hline
$R_0$ & $\mathbb{Z}$   \\
$R_1$ & $\mathbb{Z}_2$ \\
$R_2$ & $\mathbb{Z}_2$ \\
$R_3$ & $0$            \\
$R_4$ & $\mathbb{Z}  $ \\
$R_5$ & $0$            \\
$R_6$ & $0$            \\
$R_7$ & $0$            \\ \hline\hline
\end{tabular}
\end{center}
\caption{
Classifying space and the corresponding homotopy $\pi_0(C_q)$ or $\pi_0(R_q)$.
}
\label{table: homotopy}
\end{table}

In the following, we apply the above scheme to each case of AZ+$\mathcal{I}$ symmetry classes with additional chiral symmetry.

\subsection{
Application to systems of AZ+$\mathcal{I}$ classes with additional chiral symmetry
}

\subsubsection{
Class A
}
In this case, the Hamiltonian satisfies
\begin{eqnarray}
 \{H(\bm{k}),\Sigma \} &=& 0.
\end{eqnarray}
Thus, with the kinetic terms and the symmetry operator, we can consider a Clifford algebra with the following generators
\begin{eqnarray}
 \{\gamma_1,\cdots,\gamma_d,\Sigma\}.
\end{eqnarray}
Introducing the mass term, the extension problem is written as $Cl_{1+d}\to Cl_{2+d}$ where $Cl_{2+d}$ has generators
\begin{eqnarray}
\{\gamma_0,\gamma_1,\cdots,\gamma_d,\Sigma\}.
\end{eqnarray}
Thus, the corresponding classifying space is $C_{1+d}$, which results in the classification result $\pi_0(C_{1+d})$.

\subsubsection{
Class AIII
}
In this case, the Hamiltonian satisfies
\begin{subequations}
\begin{eqnarray}
 \{H(\bm{k}),\Sigma \} &=& 0, \\
 \{H(\bm{k}),\Gamma \} &=& 0.
\end{eqnarray}
\end{subequations}
In addition, $\Sigma$ and $\Gamma$ anti-commute with each other [see Eq.~(\ref{eq: comm_TP_CP_Sigma}c)].

Because the product $U=i\Gamma\Sigma$ ($U^2=1$) commute with the Hamiltonian, one can block-diagonalize the system into the $\pm 1$ sector of $U$.
The relations $\{U,\Sigma\}=\{U,\Gamma \}=0$ indicate that applying $\Sigma$ or $\Gamma$ exchanges the plus-and minus-sectors. 
In other words, no symmetry is closed for each sector.
Thus, we can classify the mass terms of the block-diagonalized Hamiltonian with the classifying space for the extension problem $Cl_d\to Cl_{1+d}$ where
$Cl_{1+d}$ has generators
\begin{eqnarray}
\{\gamma_0,\gamma_1,\cdots,\gamma_d \}.
\end{eqnarray}
Therefore, the classification result is $\pi_0(C_d)$.

\subsubsection{
Class AI and AII
}
%
In this case, the Hamiltonian satisfies
\begin{subequations}
\begin{eqnarray}
 (TP) H(\bm{k}) (TP)^{-1} &=& H(\bm{k}), \\
 \Sigma H(\bm{k}) \Sigma^{-1} &=& -H(\bm{k}),
\end{eqnarray}
with the ant-commutation relation of $TP$ and $\Sigma$ [see Eq.~(\ref{eq: comm_TP_CP_Sigma}a)].
\end{subequations}
Noticing the relation $(TP) k_j \gamma_j(TP)^{-1}=k_j\gamma_j$ for $j=1,\cdots,d$, we have $(TP)\gamma_j(TP)^{-1}=\gamma_j$.

Thus, for class AI [$(TP)^2=1$], the classifying space for the extension problem is written as $Cl_{d,3}\to Cl_{d+1,3}$ where $Cl_{d+1,3}$ is given by
\begin{eqnarray}
\{J\gamma_0,J\gamma_1,\cdots,J\gamma_d; TP, JTP,\Sigma \}.
\end{eqnarray}
Thus, the corresponding classifying space is $R_{d-1}$.

In a similar way, we can obtain the classifying space for class AII [$(TP)^2=-1$].
The extension problem is written as $Cl_{d+2,1}\to Cl_{d+3,1}$ where $Cl_{d+3,1}$ has generators
\begin{eqnarray}
\{J\gamma_0, TP, JTP ,J\gamma_1,\cdots,J\gamma_d ; \Sigma \}.
\end{eqnarray}
Thus, the corresponding classifying space is $R_{d+3}$.

Therefore, we obtain the classification results $\pi_0(R_{d+7})$ and $\pi_0(R_{d+3})$ for class AI and AII, respectively. Here we have used the Bott periodicity $\pi_0(R_{d})=\pi_0(R_{d+8})$.

\subsubsection{
Class D and C
}

In this case, the Hamiltonian satisfies
\begin{subequations}
\begin{eqnarray}
 (CP) H(\bm{k}) (CP)^{-1} &=& -H(\bm{k}), \\
 \Sigma H(\bm{k}) \Sigma^{-1} &=& -H(\bm{k}),
\end{eqnarray}
with the commutation relation of $CP$ and $\Sigma$ [see Eq.~(\ref{eq: comm_TP_CP_Sigma}b)].
\end{subequations}
Noticing the relation $(CP) k_j \gamma_j(CP)^{-1}=-k_j\gamma_j$ for $j=1,\cdots,d$, we have $(CP)\gamma_j(CP)^{-1}=-\gamma_j$.

Thus, for class D [$(CP)^2=1$], the extension problem is written as $Cl_{1,2+d} \to Cl_{1,3+d}$ where $Cl_{1,3+d}$ has generators
\begin{eqnarray}
\{J\Sigma; \gamma_0, \gamma_1,\cdots,\gamma_d,CP, JCP\}.
\end{eqnarray}
Thus, the corresponding classifying space is $R_{d+1}$.

In a similar way, we can obtain the classifying space for class C [$(CP)^2=-1$].
The extension problem is written as $Cl_{3,d} \to Cl_{3,1+d}$ where $Cl_{3,1+d}$ has generators
\begin{eqnarray}
\{CP, JCP, J\Sigma; \gamma_0, \gamma_1,\cdots,\gamma_d \}.
\end{eqnarray}
Thus, the corresponding classifying space is $R_{d-3}$.

Therefore, we obtain the classification results $\pi_0(R_{d+1})$ [$\pi_0(R_{d+5})$] for class D (C), respectively. 
Here we have used the Bott periodicity $\pi_0(R_{d})=\pi_0(R_{d+8})$.

\subsubsection{
Class BDI, DIII, CII, and CI
}
In this case, the Hamiltonian satisfies
\begin{subequations}
\begin{eqnarray}
 (TP) H(\bm{k}) (TP)^{-1} &=& H(\bm{k}), \\
 (CP) H(\bm{k}) (CP)^{-1} &=& -H(\bm{k}), \\
 \Sigma H(\bm{k}) \Sigma^{-1} &=& -H(\bm{k}).
\end{eqnarray}
with $\Sigma$ anti-commuting with $TP$ and commuting with $CP$ [see Eqs.~(\ref{eq: comm_TP_CP_Sigma}a)~and~(\ref{eq: comm_TP_CP_Sigma}b)].
\end{subequations}

For class BDI [$(TP)^2=(CP)^2=1$], the extension problem is written as $Cl_{1+d,3}\to Cl_{2+d,3}$ where $Cl_{2+d,3}$ has generators
\begin{eqnarray}
\{ JTC, J\gamma_0,J\gamma_1,\cdots, J\gamma_d;TP, JTP,\Sigma \}.
\end{eqnarray}
Thus, the corresponding classifying space is $R_{d}$.

For class DIII [$(TP)^2=-1$ and $(CP)^2=1$], the extension problem is written as $Cl_{2+d,2}\to Cl_{3+d,2}$ where $Cl_{3+d,2}$ has generators
\begin{eqnarray}
\{ TP, JTP, J\gamma_0,J\gamma_1,\cdots, J\gamma_d; JTC,\Sigma \}.
\end{eqnarray}
Thus, the corresponding classifying space is $R_{d+2}$.

For class CII [$(TP)^2=-1$ and $(CP)^2=-1$], the extension problem is written as $Cl_{3+d,1}\to Cl_{4+d,1}$ where $Cl_{4+d,1}$ has generators
\begin{eqnarray}
\{ TP, JTP, JTC, J\gamma_0,J\gamma_1,\cdots, J\gamma_d;\Sigma \}.
\end{eqnarray}
Thus, the corresponding classifying space is $R_{d+4}$.

For class CI [$(TP)^2=1$ and $(CP)^2=-1$], the extension problem is written as $Cl_{d,4}\to Cl_{1+d,4}$ where $Cl_{1+d,4}$ has generators
\begin{eqnarray}
\{ J\gamma_0,J\gamma_1,\cdots, J\gamma_d;TP, JTP, JTC,\Sigma \}. 
\end{eqnarray}
Thus, the corresponding classifying space is $R_{d-2}$.

From above calculation, we obtain the classification results; for class BDI, DIII, CII, and CI, we see $\pi_0(R_d)$, $\pi_0(R_{d+2})$, $\pi_0(R_{d+4})$, and $\pi_0(R_{d+6})$, respectively.
Here we have used the Bott periodicity $\pi_0(R_{d})=\pi_0(R_{d+8})$.

\subsection{
Application to the symmetry classes CII, C and CI with additional chiral symmetry and $CP'$-symmetry
}
\label{sec: class_Hermi_Sig_CII_C_CI_app}

\begin{table}[htb]
\begin{center}
\begin{tabular}{c c c c c c c c c c c c c} \hline\hline
             & $TP$ & $CP$ & $\Gamma$ & homotopy                       & $d=0$           & 1               &    2            & 3               & 4               & 5               & 6               &    7            \\ \hline
C            & $0$  & $-1$ & $0$              &     $\pi_0(C_{d+1})$           &   0             &  $\mathbb{Z}$   &   0             &  $\mathbb{Z}$   &   0             &  $\mathbb{Z}$   &   0             &  $\mathbb{Z}$   \\
CII${}_{+}$  & $-1$ & $-1$ & $1$              &     $\pi_0(C_{d})$             &  $\mathbb{Z}$   &   0             &  $\mathbb{Z}$   &   0             &  $\mathbb{Z}$   &   0             &  $\mathbb{Z}$   &   0             \\
CI${}_{+}$   & $1$  & $-1$ & $1$              &     $\pi_0(C_{d})$             &  $\mathbb{Z}$   &   0             &  $\mathbb{Z}$   &   0             &  $\mathbb{Z}$   &   0             &  $\mathbb{Z}$   &   0             \\
CII${}_{-}$  & $-1$ & $-1$ & $1$              &     $\pi_0(R_{d+3})$           &    0            & $\mathbb{Z}$    &    0            &    0            &    0            &  $\mathbb{Z}$   &  $\mathbb{Z}_2$ &  $\mathbb{Z}_2$ \\ 
CI${}_{-}$   & $1$  & $-1$ & $1$              &     $\pi_0(R_{d+7})$           &    0            & $\mathbb{Z}$    & $\mathbb{Z}_2$  & $\mathbb{Z}_2$  &    0            & $\mathbb{Z}$    &    0            &    0            \\ \hline
\end{tabular}
\end{center}
\caption{
Classification results for Hermitian systems of AZ+$\mathcal{I}$ symmetry classes with additional chiral symmetry and $CP'$-symmetry.
``$0$" in the second, the third, and the fourth columns denotes that the corresponding symmetry is absent.
The $\pm 1$ in the second [the third] column represents the sign of $(TP)^2=\pm 1$ [$(CP)^2=\pm 1$], respectively.
In the presence of $TP$-symmetry, the symmetry class is $\mathrm{CII}_{+}$ or $\mathrm{CI}_{+}$ ($\mathrm{CII}_{-}$ or $\mathrm{CI}_{-}$) when the additional symmetry satisfies $[CP, CP']=0$ ($\{CP, CP'\}=0$), respectively.
}
\label{table: Hermi_class_AZ+I_CII_C_CI}
\end{table}

As discussed in Sec.~\ref{sec: class_scheme_nonH}, mechanical systems, preserving particle-hole symmetry [$(CP)^2=-1$], possesses additional particle-hole symmetry $CP'$ with $(CP')^2=1$ which commutes with $\Sigma$.
Thus, we need to address the classification of the Hermitian Hamiltonian by taking into account these two types of particle-hole symmetry.

The classification results are summarized in Table~\ref{table: Hermi_class_AZ+I_CII_C_CI}.
By a proper choice of gauge, we can see that the operators $CP$ and $CP'$ satisfy the commutation or the anti-commutation relation, and that both of $CP$ and $CP'$ commute with $TP$.
The subscript of symmetry class indicates the commutation ($[CP,CP']=0$) or anti-commutation ($\{CP,CP'\}=0$) relation, e.g., for class $\mathrm{CII}_+$, the commutation relation ($[CP,CP']=0$) holds.
Here, the classification result for each case is obtained by assuming 
\begin{eqnarray}
 [CP',\Sigma]&=&0,
\end{eqnarray}
as well as Eq.~(\ref{eq: comm_TP_CP_Sigma}). We note that generic mechanical systems satisfy these relations

In the following we discuss the details.

\subsubsection{Class C}
We note that a proper choice of gauge results in $[CP,CP']=0$ in the absence of the time-reversal symmetry.
Thus, we address the classification for $[CP,CP']=0$.

In this case, the Hamiltonian satisfies the following relations
\begin{subequations}
\label{eq: summ_classC}
\begin{eqnarray}
 (CP) H(\bm{k}) (CP)^{-1} &=& -H(\bm{k}), \\
 \Sigma H(\bm{k}) \Sigma^{-1} &=& -H(\bm{k}), \\
 U H(\bm{k}) U^{-1} &=& H(\bm{k}), 
\end{eqnarray}
\end{subequations}
with $U=CPCP'$.
Here, we note that the following relations hold
\begin{subequations}
\begin{eqnarray}
{} U^2&=&-1, \\
{} [U,CP]&=&0, \\
{} [U,\Sigma]&=&0.
\end{eqnarray}
\end{subequations}
Because Eq.~(\ref{eq: summ_classC}c) holds, the Hamiltonian can be block-diagonalized with the $\pm i$-sector of $U$.
Applying the $CP$-transformation exchanges the $\pm i$-sector while the chiral symmetry is closed for each sector.

Therefore, the mass terms of the block-diagonalized Hamiltonian are classified as follows.
For each sector of $U$, the extension problem is written as $Cl_{1+d} \to Cl_{2+d}$ where $Cl_{2+d}$ has generators
\begin{eqnarray}
\{ \Sigma, \gamma_0, \gamma_1,\cdots,\gamma_d\}.
\end{eqnarray}
Thus the classifying space for the above extension problem is $C_{1+d}$.
Noticing that $CP$-transformation maps the Hamiltonian of $-i$-sector to that of $+i$-sector, we obtain the classification result $\pi_0(C_{1+d})$.

\subsubsection{Class $\mathrm{CII}_{+}$}

For symmetry class $\mathrm{CII}_{+}$ where $[CP,CP']=0$ holds, the following symmetry constraints are satisfied.

\begin{subequations}
\label{eq: summ_classCIIandCI}
\begin{eqnarray}
 (TP) H(\bm{k}) (TP)^{-1} &=& -H(\bm{k}), \\
 \Gamma H(\bm{k}) \Gamma^{-1} &=& -H(\bm{k}), \\
 \Sigma H(\bm{k}) \Sigma^{-1} &=& -H(\bm{k}), \\
 U H(\bm{k}) U^{-1} &=& H(\bm{k}), 
\end{eqnarray}
\end{subequations}
with $U=CPCP'$ ($U^2=-1$) and $\Gamma=TPCP$.

Defining $V=\Gamma\Sigma$ with $V^2=-1$ [see Eq.~(\ref{eq: comm_TP_CP_Sigma})], we have the following set of the operators.
\begin{subequations}
\label{eq: U_V_CP_comm_CP'}
\begin{eqnarray}
{}[V,U] &=& 0, \\
{}\{V,TP\} &=& 0, \\
{}\{V,\Gamma\} &=& 0, \\
{}[U,TP] &=& 0, \\
{}[U,\Gamma] &=& 0, \\
{}[\Gamma,TP] &=& 0,
\end{eqnarray}
\end{subequations}
where unitary matrices $U$ and $V$ commute with the Hamiltonian.

Thus, the Hamiltonian can be block-diagonalized with $U$ and $V$.
Each sector is labeled with the set of eigenvalues $(u,v)$ where $u=\pm i$ ($v=\pm i$) denotes eigenvalue of $U$ ($V$), respectively.
Eqs.~(\ref{eq: U_V_CP_comm_CP'}b)~and~(\ref{eq: U_V_CP_comm_CP'}d) indicate that applying the operator $TP$ maps $(u,v)$-sector to $(-u,v)$-sector. 
Eqs.~(\ref{eq: U_V_CP_comm_CP'}c)~and~(\ref{eq: U_V_CP_comm_CP'}e) indicate that applying the operator $\Gamma$ maps $(u,v)$-sector to $(u,-v)$-sector.
These facts indicate that no symmetry is closed for each sector.

Therefore, the mass terms of the block-diagonalized Hamiltonian are classified as follows.
For each sector, the extension problem is written as $Cl_{d} \to Cl_{1+d}$ where $Cl_{1+d}$ has generators
\begin{eqnarray}
\{ \gamma_0, \gamma_1,\cdots,\gamma_d\}. 
\end{eqnarray}
Thus, the classifying space for the above extension problem is $C_{d}$.
Noticing that applying $TP$ or $\Gamma$ maps the Hamiltonian of $(i,i)$-sector to that of the other sectors, we obtain the classification result $\pi_0(C_{d})$.

\subsubsection{
Class $\mathrm{CI}_{+}$
}
We consider the case where $(TP)^2=1$ and $\{CP,CP'\}=0$ holds.

Defining $\Gamma=iTPCP$ ($\Gamma^2=1$), we have Eq.~(\ref{eq: summ_classCIIandCI}).
We note that 
\begin{subequations}
\label{eq: symm_CI+}
\begin{eqnarray}
 [V,TP]&=&0, \\
 \{\Gamma,TP\}&=&0, 
\end{eqnarray}
\end{subequations}
hold instead of Eqs.~(\ref{eq: U_V_CP_comm_CP'}b)~and~(\ref{eq: U_V_CP_comm_CP'}f). The other relations in Eq.~(\ref{eq: U_V_CP_comm_CP'}) hold.

In this case, the Hamiltonian can be block-diagonalized for $(u,v)$-sectors where $u=\pm i$ ($v=\pm i$) denotes eigenvalue of $U$ ($V$), respectively.
Eqs.~(\ref{eq: U_V_CP_comm_CP'}d)~and~(\ref{eq: symm_CI+}a) indicate that applying the operator $TP$ flips sign both of $u$ and $v$.
Eqs.~(\ref{eq: U_V_CP_comm_CP'}c)~and~( \ref{eq: U_V_CP_comm_CP'}e) indicate that applying the operator $\Gamma$ flips sign of $v$.

Therefore, the mass terms of the block-diagonalized Hamiltonian are classified as follows.
For each sector, the extension problem is written as $Cl_{d} \to Cl_{1+d}$ where $Cl_{1+d}$ has generators
\begin{eqnarray}
\{ \gamma_0, \gamma_1,\cdots,\gamma_d\}.
\end{eqnarray}
Thus, the classifying space for the above extension problem is $C_{d}$.
Noticing that applying $TP$ or $\Gamma$ maps the Hamiltonian of $(i,i)$-sector to that of the other sectors, we obtain the classification result $\pi_0(C_{d})$.

\subsubsection{
Class $\mathrm{CII}_{-}$
}
We consider the case where the Hamiltonian satisfies Eq.~(\ref{eq: summ_classCIIandCI}) with $U=CPCP'$ ($U^2=1$) and $\{CP,CP'\}=0$.

In this case, the operators satisfy the following relations
\begin{subequations}
\label{eq: U_V_CP_an-comm_CP'}
\begin{eqnarray}
{}[U,TP] &=& 0, \\
{}\{U,\Gamma\} &=& 0, \\
{}\{U,V\} &=& 0,\\
{}\{V,TP\} &=& 0,\\
{}\{V,\Gamma\} &=& 0,
\end{eqnarray}
\end{subequations}
with $\Gamma=TPCP$ and $V=\Gamma\Sigma$ ($V^2=-1$).

Thus, the Hamiltonian can be block-diagonalized with $U$.
We can see that applying $\Gamma$ or $V$ exchanges the plus-and the minus-sectors, while the $TP$-symmetry is closed for each sector [see Eq.~(\ref{eq: U_V_CP_an-comm_CP'})]. 
Namely, the block-diagonalized Hamiltonian preserves the symmetry of $\Sigma$ and $TP$.

Therefore, the mass terms of the block-diagonalized Hamiltonian are classified as follows.
For each sector, the extension problem is written as $Cl_{2+d,1}\to Cl_{3+d,1}$, where $Cl_{3+d,1}$ has generators
\begin{eqnarray}
\{TP, JTP,J\gamma_0,J\gamma_1,\cdots,J\gamma_d;\Sigma \}.
\end{eqnarray}
Thus, the classifying space for the above extension problem is $R_{d+3}$.
Noticing that applying $\Gamma$ maps the Hamiltonian of $i$-sector to that of the other sector, we obtain the classification result $\pi_0(R_{d+3})$.

\subsubsection{
Class $\mathrm{CI}_{-}$
}
We consider the case where $(TP)^2=1$ and $\{CP,CP'\}=0$ hold. 
With $\Gamma=iTPTC$ and $U=CPCP'$ ($U^2=1$), we can see that the Hamiltonian satisfies Eq.~(\ref{eq: summ_classCIIandCI}).

In this case, we obtain the classification results in a similar way to the previous case.
First, we note that 
\begin{eqnarray}
\label{eq: symm_CI-}
[V,TP]=0,
\end{eqnarray}
holds instead of Eq.~(\ref{eq: U_V_CP_an-comm_CP'}d). The other relations of Eq.~(\ref{eq: U_V_CP_an-comm_CP'}) holds.
Thus, block-diagonalizing the Hamiltonian with $U$, we can see that applying $\Gamma$ or $V$ exchanges the plus and the minus sectors, while  the $TP$-symmetry is closed for each sector [see Eq.~(\ref{eq: U_V_CP_an-comm_CP'})]. 
Namely, the block-diagonalized Hamiltonian preserves the symmetry of $\Sigma$ and $TP$.

Therefore, the mass terms of the block-diagonalized Hamiltonian are classified as follows.
For each sector, the extension problem is written as $Cl_{d,3}\to Cl_{1+d,3}$, where $Cl_{1+d,3}$ has generators
\begin{eqnarray}
\{J\gamma_0,J\gamma_1,\cdots,J\gamma_d; TP, JTP, \Sigma \}.
\end{eqnarray}
Thus, the classifying space for the above extension problem is $R_{d+7}$.
Noticing that applying $\Gamma$ maps the Hamiltonian of the plus-sector to that of the other sector, we obtain the classification result $\pi_0(R_{d+7})$.

\end{document}